\documentclass[useAMS,usenatbib]{mn2e}
\usepackage{epsfig}
\usepackage{amsmath}
\usepackage{rotating}

\newcommand{\be}{\begin{equation}}
\newcommand{\beq}{\begin{equation}}
\newcommand{\ba}{\begin{eqnarray}}
\newcommand{\ee}{\end{equation}}
\newcommand{\eeq}{\end{equation}}
\newcommand{\ea}{\end{eqnarray}}

\newcommand{\hs}{\hspace{1mm}}

\newcommand{\apj}{ApJ}
\newcommand{\aap}{A\&A}
\newcommand{\apjl}{ApJL}
\newcommand{\mnras}{MNRAS}
\newcommand{\aj}{AJ}
\newcommand{\apjs}{ApJS}
\newcommand{\nat}{{\it Nature}}
\newcommand{\araa}{ARA\&A}

\newcommand\ion[2]{#1$\;${\scshape{#2}}}

\def\lsim{~\rlap{$<$}{\lower 1.0ex\hbox{$\sim$}}}

\def\gsim{~\rlap{$>$}{\lower 1.0ex\hbox{$\sim$}}}

\title[Ly$\alpha$ RT through Large-Scale Clumpy Outflows]{Line Transfer through Clumpy, Large-Scale Outflows: Ly$\alpha$ Absorption and Halos around Starforming Galaxies}

\author[Dem Style]{Dem$^{1}$\thanks{E-mail:mdijkstr@cfa.harvard.edu}, }

\author[Dijkstra \& Kramer]{Mark Dijkstra$^{1}$\thanks{E-mail:dijkstra@mpa-garching.mpg.de} and Roban Kramer$^{2}$\\
$^1$Max-Planck Institut fuer Astrophysik, Karl-Schwarzschild-Str. 1, 85741 Garching, Germany\\
$^{2}$ Institute for Astronomy, ETH Zurich, Wolfgang-Pauli-Strasse 27, CH-8093 Zurich, Switzerland}

\begin{document}

\date{\today}
\pagerange{\pageref{firstpage}--\pageref{lastpage}} \pubyear{2006}

\maketitle

\label{firstpage}
\begin{abstract}
We present constrained radiative transfer calculations of Ly$\alpha$
photons propagating through clumpy, dusty, large scale outflows, and
explore whether we can quantitatively explain the Ly$\alpha$ halos
that have been observed around Lyman Break Galaxies.
 We construct phenomenological models of large-scale outflows which consist of cold clumps that are in pressure
 equilibrium with a constant--velocity hot wind. First we consider
 models in which the cold clumps are distributed symmetrically around
 the galaxy, and in which the clumps undergo a continuous acceleration in its `circumgalactic' medium (CGM). We constrain the properties of the cold clumps (radius, velocity, \ion{H}{I} column density, \& number density)
 by matching the observed Ly$\alpha$ absorption strength of
 the CGM in the spectra of background
 galaxies. We then insert a Ly$\alpha$ source in the center of this
 clumpy outflow, which consists of $10^{5-6}$ clumps,
 and compute observable properties of the scattered Ly$\alpha$
 photons. In these models, the scattered radiation forms halos that are significantly more concentrated than is observed. In order to simultaneously reproduce the observed
 Ly$\alpha$ absorption line strengths and the Ly$\alpha$ halos, we
 require -- preferably bipolar -- outflows in which the clumps decelerate after their initial acceleration. This
 deceleration is predicted naturally in `momentum--driven' wind models
 of clumpy outflows.  In models that simultaneously fit the absorption
 and emission line data, the predicted linear polarization is $\sim
 30-40\%$ at a surface brightness contour of $S=10^{-18}$ erg s$^{-1}$
 cm$^{-2}$ arcsec$^{-2}$. Our work illustrates clearly that Ly$\alpha$
 emission line halos around starforming galaxies provide valuable
 constraints on the cold gas distribution \& kinematics in their
 circumgalactic medium, and that these constraints complement those
 obtained from absorption line studies alone.
 \end{abstract}

 
\section{Introduction}
\label{sec:intro}

Deep narrowband imaging has revealed that star forming galaxies are
surrounded by large ($R\sim 100$ kpc) Ly$\alpha$ halos \citep[][see
  also Fynbo et al. 1999, Rauch et
  al. 2008]{Hayashino04,Steidel11,Matsuda12}. The origin of these halos is still
disputed, but they likely encode valuable information on both the
distribution and kinematics of cold gas around galaxies. Understanding
this so-called `circumgalactic' medium (CGM) is vital to our
understanding of galaxy formation and evolution.

 \citet{ZZex} attribute the presence of extended Ly$\alpha$ halos to
 resonant scattering of Ly$\alpha$ photons in the CGM. In this model,
 the gravitational potential well of the dark matter halo that is
 hosting the galaxy gives rise to inflows of overdense, ionized,
 circumgalactic gas. This infalling, overdense gas contains residual
 \ion{H}{I} that is opaque to Ly$\alpha$ radiation \citep[also
   see][]{BL03,Santos04,IGM,Iliev08,Laursen11}, and can scatter a
 significant fraction of the Ly$\alpha$ flux that escapes from
 galaxies into diffuse halos.

However, observations of the gas in the circum galactic medium of
Lyman Break Galaxies (LBGs) indicate that the `cold' (i.e. $T\sim
10^4-10^5$ K) gas is almost exclusively outflowing: low ionization
metal absorption lines are typically blue-shifted relative to the
galaxies' systematic redshift, and the covering factor of these
blueshifted absorption lines are large
\citep{Steidel10}. \citet{Steidel10} argue that these outflows --which
are not present in the the simulations by \citet{ZZex}-- play an
important role in the Ly$\alpha$ transport problem, and scattering
through outflows can explain ({\it i}) in particular the extended red
wing of the Ly$\alpha$ spectral shape of the Ly$\alpha$ line that is
observed in their galaxies, and ({\it ii}) the observed Ly$\alpha$
halos around their galaxies \citep{Steidel11}. \citet{Steidel10}
provide a simple model for the Ly$\alpha$ scattering process in which
a spherically symmetric distribution of outflowing clumps surrounds
each galaxy. The `covering factor' of clumps on the sky decreases, and
their outflow velocity increases with distance from the galaxy. 

 \citet{Steidel10} treat the scattering of Ly$\alpha$ photons in only
 an approximate way, and it is not clear whether a full radiative
 transfer calculation would yield similar results.  Given the simple
 geometry of the clumpy outflow model that \citet{Steidel10} propose,
 it is straightforward to treat the scattering process itself more
 accurately, and to investigate whether their model can indeed
 quantitatively reproduce the observed Ly$\alpha$ halos. It is
 important to test the scattering model, and to investigate if other
 processes need to be invoked to explain the presence of extended
 Ly$\alpha$ halos around (all) star forming galaxies. These other
 processes include for example resonant scattering in the ionized
 inflowing CGM \citep{ZZex}, and spatially extended Ly$\alpha$ {\it
   emission} (as opposed to scattering) from supernova driven outflows
 \citep{Mori04}, or from the cold gas streams that provide galaxies
 with their gas in cosmological hydrodynamical simulations
 \citep[e.g.][]{Fardal01,Furlanetto05,DL09,Goerdt10,FG10,RB12}. Indeed, recent work has indicated
 that so-called `cold-flows' may reproduce the observed Ly$\alpha$
 absorption line strengths in the CGM of simulated LBGs quite well
 \citep{Fu11}, which raises the question to whether the Ly$\alpha$
 halos are also related to cold streams\footnote{`Gravitational
   heating' of cold flow gas in dark matter halos of mass $M_{\rm
     halo}\sim 10^{12} M_{\odot}$, the approximate mass of the dark
   matter halo masses associated with LBGs \citep[e.g.][and references
     therein]{Rudie12}, can give rise to Ly$\alpha$ luminosities of
   order $L\lsim 10^{42}$ erg s$^{-1}$ \citep{H00,DL09,Goerdt10,FG10},
   which is an order of magnitude fainter than the observed
   luminosities of Ly$\alpha$ halos around LBGs. This rules out the
   possibility that gravitationally heated cold flows contributed
   significantly to the observed luminosity of Ly$\alpha$ halos
   \citep[but see][]{RB12}. However, it is possible to increase the
   Ly$\alpha$ emissivity of cold stream gas if sources embedded in the
   stream photoionize the gas in the streams
   \citep[e.g.][]{DL09,Latif11}.}.

The goal of this paper is simple: to test whether scattering through a
large-scale clumpy, possibly dusty, outflow can give rise to spatially
extended Ly$\alpha$ halos around star forming galaxies, and
importantly, whether such models are consistent with the observed
Ly$\alpha$ absorption strength of the circumgalactic medium as
measured in the spectra of background galaxies (as in Steidel et
al. 2010). In order to properly model the scattering process, we modify the
Monte-Carlo radiative transfer code for Ly$\alpha$ propagation from
\citet{D06}, so that it works on arbitrary distributions of dusty clumps. 

Having such a code is very useful, as understanding Ly$\alpha$ transfer through (clumpy, dusty) outflows is
relevant in a wider range of astrophysical contexts: Firstly,
Ly$\alpha$ radiative transfer through outflows affects the imprint
that reionization leaves on the visibility of the Ly$\alpha$ emission
line from high redshift galaxies
\citep{Santos04,DW10,D11,DF11}. Secondly, scattering through outflows
strongly affects the large scale clustering of Ly$\alpha$ selected
galaxies in the post-reionization epoch \citep{ZZclus,WD11}, which is
directly relevant for e.g. the HETDEX\footnote{http://hetdex.org/}
dark energy experiment \citep{Hill04}. Finally, Ly$\alpha$ line
transfer through clumpy, dusty (not necessarily outflowing) media is
of interest because it can strongly boost the fraction of Ly$\alpha$
photons that can escape from the interstellar medium of a galaxy,
possibly even enhancing the equivalent width of the line as it emerges
from the galaxy  \citep[][but also see Scarlata et al. 2009, Kornei et
  al. 2010]{Neufeld91,HS99,Ho06,Finkelstein09}. 
  
This paper is organized as follows: In \S~\ref{sec:clump1} we
introduce the basic quantities that describe a general clump
distribution. In \S~\ref{sec:clump2} we describe our model for the
cold clumps embedded within a hot, large scale outflow, and how we
constrain the parameters of this model by matching to the absorption
line data of Steidel et al. (2010). This procedure fixes the
properties of the scattering medium. We describe how we perform
Ly$\alpha$ radiative transfer calculations through this medium in
\S~\ref{sec:lyart}.  We present the main results of our calculations,
and explore how these depend on our assumed initial line profile and
dust content of the clumps in \S~\ref{sec:results}. In
\S~\ref{sec:vel} we explore more realistic velocity profiles of the
cold clumps. We discuss model uncertainties and broader implications
of our work in \S~\ref{sec:discuss} before presenting our main
conclusions in \S~\ref{sec:conc}.

In this paper we express frequency $\nu$ in terms of the dimensionless
variable $x\equiv (\nu-\nu_{\alpha})/\Delta \nu_{\alpha}$. Here,
$\nu_{\alpha}=2.46 \times 10^{15}$ Hz denotes the frequency
corresponding the Ly$\alpha$ resonance (similarly
$\lambda_{\alpha}=1215.67$ \AA\hs corresponds to wavelength of this
transition), and $\Delta \nu_{\alpha} \equiv \nu_{\alpha}\sqrt{2kT/m_p
  c^2}\equiv \nu_{\alpha}v_{\rm th}/c$. Here, $T$ denotes the
temperature of the gas that is scattering the Ly$\alpha$
radiation. Table~\ref{table:symbols} gives an overview of symbols used
in this paper. The cosmological parameters used throughout our
discussion are $(\Omega_m,\Omega_{\Lambda},h)=(0.3,0.7,0.7)$, which is
consistent with  \citet{Komatsu09}.

\section{Some General Clump Statistics}
\label{sec:clump1}

\begin{table}
\caption{Summary of symbols used throughout this paper.}  \centering.
\begin{tabular}{l l}
\hline\hline \multicolumn{2}{|c|}{Symbols describing clump
  properties.}\\ \hline\hline $r$ & physical separation of a clump
from the source  (kpc) \\ $b$ & `impact' parameter (kpc): \\ $n_c$ &
number density of clumps (in kpc$^{-3}$) \\ $n_{\rm H}$ & number
density of H-nuclei (i.e. protons, in cm$^{-3}$)\\ $n_{\rm HI}$ &
number density of neutral hydrogen atoms (in cm$^{-3}$)\\ $R_c$ &
radius of the clump \\ $\sigma_c$ & cross-sectional area of the clump
(in kpc$^2$):\\ & $\sigma_c=\pi R^2_c$\\ $m_{\rm c}$ & clump mass (in $M_{\odot}$)\\
$N_{\rm HI}$ & column density
of \ion{H}{I} through the clump,\\ & weighted by cross-sectional area
(in cm$^{-2}$):\\ & $N_{\rm HI}=4 R_c n_{\rm HI}/3$\\ $f_{\rm c}$ &
covering factor (in kpc$^{-1}$):\\ & $f_{\rm c}=n_c\sigma_c$\\ $T_{\rm
  c}$ & gas temperature in the cold clumps \\ $v_{\rm th}$ & `thermal'
velocity of \ion{H}{I} in cold clumps (km s$^{-1}$) \\ & $v_{\rm
  th}=\sqrt{2kT_{\rm c}/m_{\rm p}}$\\ \hline\hline
\multicolumn{2}{|c|}{Symbols used for Ly$\alpha$ radiative
  transfer.}\\ \hline\hline $\nu_{\alpha}$ & Ly$\alpha$ resonance
frequency\\ & $\nu_{\alpha}=2.46 \times 10^{15}$ Hz\\ $\Delta
\nu_{\alpha}$ & thermal line broadening (Hz)\\ & $\Delta \nu_{\alpha}
\equiv \nu_{\alpha}v_{\rm th}/c$\\ $x$ & dimensionless photon
frequency \\ & $x\equiv (\nu-\nu_{\alpha})/\Delta
\nu_{\alpha}$\\ $\sigma_0$ & Ly$\alpha$ absorption cross-section at
line center\\ & $\sigma_0=5.88 \times 10^{-14}(T_{\rm c}/10^{4}\hs{\rm
  K})^{-1/2}$ cm$^{2}$\\ $\sigma_{\alpha}(x)$ & Ly$\alpha$ absorption
cross-section at frequency $x$ (cm$^{2}$) \\ &
$\sigma_{\alpha}(x)=\sigma_0\phi(x)$\\ $\phi(x)$ & Voigt function at
frequency $x$. \\ &  We adopt the normalization such that
$\phi(x=0)=1$ \\ &  , and therefore that $\int
\phi(x)dx=\sqrt{\pi}$.\\ $a_{\rm v}$ & Voigt parameter\\ & $a_{\rm
  v}=4.7 \times 10^{-4}(T_c/10^{4}\hs{\rm K})^{-1/2}$ \\ ${\bf k}_{\rm
  in/out}$ & Unit vector that denotes the propagation direction\\ &
of the photons before/after scattering.\\ ${\bf e}_{\rm in/out}$ &
Unit vector that denotes the electric vector\\ & of the photons
before/after scattering.\\ $\mu$ & Cosine of the scattering angle\\ &
$\mu ={\bf k}_{\rm in} \cdot {\bf k}_{\rm out}$\\ $P(\mu)$ &Scattering
phase function\\ & We adopt the normalization $\int _{-1}^{1}d\mu\hs
P(\mu)=4 \pi$\\ \hline\hline
\end{tabular}
\label{table:symbols}
\end{table}
We denote the number density of clumps at a separation $r$ from the
source (located at $r=0$) by $n_c(r)$, and their outflow velocity by
$v_c(r)$. We denote the radius of a clump by $R_c(r)$. The number
density of neutral hydrogen atoms inside clumps is denoted by $n_{\rm
  HI}(r)$. The {\it average} HI-column density of the cold clumps is
denoted by $N_{\rm HI}(r)$ and is given\footnote{The average column
  density is given by $N_{\rm HI}(r)=\frac{1}{\pi
    R^2_c(r)}\int_0^{R_c(r)}dy\hs 2\pi y \hat{N}_{\rm HI}(y)$, where
  $\hat{N}_{\rm HI}(y)$ denotes the total HI column density at impact
  parameter $y$ through the clump, and is given by $\hat{N}_{\rm
    HI}(y)=2\sqrt{R^2_c-y^2}n_{\rm HI}$. We can evaluate the integral
  analytically and obtain $N_{\rm HI}(r)=4 n_{\rm HI}(r)R_c(r)/3$.} by
$N_{\rm HI}(r)\equiv 4 n_{\rm HI}(r)R_c(r)/3$.  Finally, the {\it
  covering factor} $f_c(r)$ is given by $f_c(r) = n_c(r) \sigma_c(r)$,
where $\sigma_c(r) = \pi R_c(r)^2$ denotes the geometric cross-section
of a clump of radius $R_c(r)$. The covering factor $f_c(r)$ thus has
units of length$^{-1}$, and plays a role that is analogous to opacity
$\kappa(r)$ in a homogeneous medium. We caution the reader that our
definition of covering factor differs from that adopted by
\citet{Ho06}, who defined the covering factor to be the mean {\it
  total} number of clumps encountered along a random line of sight
(which we denote by $\mathcal{N}_{\rm clump}$, see below), and which
is therefore analogous to optical depth in a homogeneous
medium. Table~\ref{table:symbols} summarizes the symbols that we use
to describe the clump properties.

Many useful properties of the clumps can expressed in terms of this
covering factor:

\begin{itemize}

\item The `mean free path' between clumps $\lambda_{\rm
  c}(r)=1/[n_c(r) \sigma_c(r)]=1/f_c(r)$. 

\item Hence, the mean number of clumps that a random sightline through
  the distribution of clumps
\begin{equation}
\mathcal{N}_{\rm clump}(b)=2 \int_b^{r_{\rm
    max}}\frac{rdr}{\sqrt{r^2-b^2}}f_c(r),
\end{equation}  where $b$ denotes the `impact parameter' of the sightline, which is the perpendicular distance of the sightline to the origin $r=0$. We integrate out to radius $r_{\rm max}=250$ kpc, which corresponds to the radius where the observed absorption vanishes (this corresponds to $R_{\rm eff}$ in Steidel et al. 2010).

\item The `shell covering factor' $F_c(r)$ which denotes the fraction
  of the area of a spherical shell at radius $r$ that is embedded
  within clumps (denoted by $C_f(r)$ by Martin \& Bouch\'{e} 2009) is
  given by

\begin{eqnarray}
F_c(r)=\int_0^\infty dr'\hs n_c(r')\sigma_c(r|r')= \\ \nonumber
\approx \pi \int_{r-R_c(r)}^{r+R_c(r)}dr'\hs
n_c(r')[R^2_c(r')-(r-r')^2]\approx \\ \nonumber \frac{4}{3}n_c(r)\pi
R^3_c(r)=\frac{4}{3}f_c(r)R_c(r)=f_v(r),
\end{eqnarray} where $\sigma_c(r|r')$ denotes the area of the clump whose center lies at radius $r'$ on the sphere of radius $r$. The approximation on the second line is only true when $dR_c(r)/dr \ll 1$, which is generally true in this paper. The quantity $f_v(r)$ denotes the volume filling factor of clumps at radius $r$, and is given by $f_v(r)=n_c(r)\frac{4}{3}\pi R^3_c(r)$.

\end{itemize}

If the total mass outflow rate is given by $\dot{M}_{\rm out}$, then
the total mass flux through each radial shell is given by
$\dot{M}_{\rm out}/4 \pi r^2$. The mass density\footnote{The analogy
  with radiation is illuminating: suppose a central source of
  radiation (instead of mass) has a luminosity $L$. Then the energy
  flux through radial shell $r$ is $s=L/4 \pi r^2$. The energy density
  in the radiation field is $u = L/4 c \pi r^2$  (e.g. Rybicki \& Lightman 1997, also see Fig~1 of
  Dijkstra \& Loeb 2008b).} in clumps at radius $r$ is then $\rho_c(r)
= \dot{M}_{\rm out}/4 \pi r^2 v_c(r)$. If the clumps have a constant
mass, then the number density of clumps is $n_{c}(r) = \dot{N}_{\rm
  c}/4 \pi r^2 v_c(r)$, where $\dot{N}_{\rm c}$ is the total rate at
which clumps are ejected.  For a constant $\dot{N}_c$, the radial
dependence of $\sigma_c(r)=f_c(r)/n_c(r) \propto r^2f_c(r)v_c(r)$,
i.e. $R_c(r) \propto r f_c^{1/2}(r)v_c^{1/2}(r)$. The total number of
clumps is given by $N_{\rm clump}=\int_0^{r_{\rm max}} dr\hs 4 \pi r^2
n_c(r)$.

\section{Modeling Clumpy Outflows around LBGs}
\label{sec:clump2}

The theory of large-scale outflows around star forming galaxies is
extremely complex. Furthermore, the kinematics and distribution of
cold gas in the outflow is particularly uncertain. Energy and momentum
deposition by radiation, supernova explosions and cosmic rays create
hot overpressurized bubbles, which sweep up the surrounding ISM into
dense, cold shells of gas \citep[see e.g. the introduction of][and
  references therein]{DS08,CK09}. These `supershells' accelerate as
they break out of interstellar medium into the lower density CGM,
making them subject to hydrodynamical (Rayleigh-Taylor)
instabilities. The overall acceleration -- and hence the velocity that
the cold gas can reach -- depends sensitively on when the cold gaseous
shells fragment: after fragmentation, the hot gas can expand freely
through the fragmented shell, which reduces the outward pressure on
the cold gas \citep[e.g.][and references therein]{Cooper,Fujita09}. 

\citet{Fujita09} have shown that a spatial resolution of $\sim 0.1$ pc
is required to resolve the hydrodynamical instabilities, and highlight
the physics that likely affects the detailed properties of the cold
gas in the outflow. For example: magnetic fields may prevent
fragmentation of the gas shells and thus allow the cold gas to reach
larger velocities; thermal conduction may further stabilize the cold
shells. On the other hand, photoionization may heat the cold clumps to
higher temperatures, which reduces the density contrast with the hot
wind, which may in turn enhance the fragmentation of the cold shells. 

Regardless of these complex model details, we expect cold fragments of
gas to be entrained within a hot wind, and that these cold clumps have
outflow velocities that are less than or equal to the hot wind outflow
velocity. The fate of these cold clumps is unclear: \citet{Klein94}
showed that the cold clumps are destroyed on a short time scale as a
result of hydrodynamical instabilities at the cold cloud--hot wind
interface. Recent studies have shown that including thermal conduction
and magnetic fields in the calculations may significantly enhance the
survival probability of the cold clumps \citep[see][and references
  therein]{Cooper,Fujita09}. Additionally, new cold clumps might form out
from thermal instabilities in the hot wind (or hot halo gas,
e.g. Joung et al. 2012). These two effects (cloud destruction and
formation) introduce further uncertainties to the spatial distribution
of cold clumps.
  
The previous illustrates clearly that ab initio modeling of cold gas
in a large scale galactic outflow is extremely complex. We therefore
adopt a simple phenomenological model for the cold clouds in the
outflow as in \citet{MB09}. Following Steidel et al. (2010), we assume that these clouds are distributed symmetrically around the galaxy.
This simple model contains parameters,
which we will constrain by matching the absorption line data of
\citet{Steidel10}. 

\subsection{The Model}
\label{sec:cmod}
The total mass outflow rate $\dot{M}_{\rm out} \equiv \eta \times$SFR,
where `$\eta$' denotes the `mass-loading' factor
\citep[e.g.][]{OD06,Dave11}.  We assume SFR=34 $M_{\odot}$ yr$^{-1}$
which corresponds to the median UV-based dust-corrected SFR of all 92
continuum selected galaxies that were used to create the stacked
Ly$\alpha$ image that revealed the Ly$\alpha$ halo surrounding
them\footnote{The galaxies that were used to compile the stacked
  Ly$\alpha$ image are not the same galaxies for which the CGM was
  probed with background galaxies (see \S~\ref{sec:caveats}). However,
  both sets of galaxies were selected in a very similar way and have
  very similar physical properties. For example, the median star
  formation rate of the `CGM galaxies' was
  SFR$=30\hs M_{\odot}$ yr$^{-1}$ \citep{Erb}, which is practically
  indistinguishable from the value that we have adopted.}. We assume
the outflow consists of a `cold' component' which is embedded in a
`hot' component, both of which are in pressure equilibrium.  The total
cold [hot] mass outflow rate is denoted by $\dot{M}_{\rm c}$
[$\dot{M}_{\rm h}$]. We further assume that the clumps all have the same mass
(and radius) when they are driven out, and that their mass does not
change while they propagate out. Under this assumption the mass of an
individual clump, $m_{\rm c}$, relates to the total number of clumps in our
Monte-Carlo simulation as:

\begin{equation}
m_{\rm c}=\frac{\dot{M}_{\rm c}}{N_{\rm clump}}\int_0^{r_{\rm
    max}}\frac{dr}{v_{\rm c}(r)}\equiv \frac{\eta_{\rm c}{\rm
    SFR}}{N_{\rm clump}}\int_0^{r_{\rm max}}\frac{dr}{v_{\rm c}(r)},
\label{eq:mclump}
\end{equation} where we introduced the cold-gas mass loading factor $\eta_{\rm c}\equiv \dot{M}_{\rm c}/{\rm SFR}$. For a given $N_{\rm clump}$ we need to know the velocity profile $v_c(r)$ to compute the mass in the cold clumps. In order to compute the HI gas density inside the clump, we need to know pressure in the hot wind.

Following \citet{MB09} we assume a steady-state constant velocity hot
wind, for which mass conservation implies $\rho_{\rm
  h}(r)=\dot{M}_{\rm h}/(4 \pi r^2 v_{\rm h})$, in which $v_{\rm h}$
denotes the outflow velocity of the hot wind. The number density of
particles in the hot-wind is $n_{\rm h}(r) \approx \rho_{\rm
  h}(r)/m_p$, where we assumed that the mean molecular weight
$\mu_{\rm h}\approx 1$. Both components obey
$p(r)n^{-\gamma}(r)=$constant, where $\gamma$ denotes the gas'
adiabatic index, and their gas pressure is given by $p(r)=n(r) k_{\rm
  B}T(r)$. Assuming pressure equilibrium between the hot and cold gas,
we find
\begin{eqnarray}
T_{\rm h}(r)=T_{\rm h,0}\Big{(} \frac{r}{r_{\rm
    min}}\Big{)}^{2-2\gamma_{\rm h}},\\ \nonumber T_{\rm c}(r)=T_{\rm
  c,0}\Big{(}\frac{r}{r_{\rm min}}\Big{)}^{-2\gamma_{\rm
    h}+2\gamma_{\rm h}/\gamma_{\rm c}}, \\ \nonumber n_{\rm
  c,H}=\frac{T_{\rm h,0}}{T_{\rm c,0}}\frac{\dot{M}_{\rm h}}{m_p \hs
  4\pi r_{\rm min}^2 v_{\rm h}}\Big{(} \frac{r}{r_{\rm
    min}}\Big{)}^{-2\gamma_{\rm h}/\gamma_{\rm c}}
\end{eqnarray} Here, $\gamma_{\rm c}$ [$\gamma_{\rm h}$] denotes the adiabatic index of the cold [hot]gas, and $T_{\rm c,0}$ [$T_{\rm h,0}$] denotes the temperature of the cold [hot] gas at some reference radius $r_{\rm min}$, which denotes the `launch' or `break--out' radius (as in Steidel et al. 2010). We further assume that the (constant) velocity of the hot wind is related to the temperature of the hot gas at the break out radius as  $v_{\rm h}\approx 940 (T_{\rm h,0}/10^7\hs{\rm K})^{0.5}$ km s$^{-1}$ \citep{MB09}. If we substitute some `typical' values the we find 

\begin{equation}
n_{\rm c,H}\approx 36\Big{(} \frac{T^{1/2}_{\rm h,7}\dot{M}_{\rm
    h,10}}{T_{\rm c,4}r^2_{\rm min,1}}\Big{)}\Big{(} \frac{r}{r_{\rm
    min}}\Big{)}^{-2\gamma_{\rm h}/\gamma_{\rm c}}\hs{\rm cm}^{-3},
\end{equation} where $T_{\rm h,7}\equiv T_{\rm h,0}/10^7\hs$K, $T_{\rm c,4}\equiv T_{\rm c,0}/10^4\hs$K,  $r_{\rm min,1}=r_{\rm min}/{\rm kpc}$, and $\dot{M}_{\rm h,10}\equiv\dot{M}_{\rm h} /[10 \hs {\rm M}_{\odot} \hs {\rm yr}^{-1}]$. The number density $n_{\rm c,H}$ refers to the total number density of hydrogen nuclei in the clump. When the clump self-shields, we expect all of the hydrogen to be neutral. Recent hydrodynamical simulations of cosmological volumes indicate that a decent approximation to the full radiative transfer of ionizing radiation is obtained by assuming that gas self-shields when the number density exceeds some threshold value of $n_{\rm crit} \gsim 6 \times 10^{-3}$ cm$^{-3}$ \citep[e.g.][]{Nagamine10}. In our model, we will assume that the number density of neutral hydrogen atoms, $n_{\rm HI,c}(r)=n_{\rm c,H}$ for $n_{\rm c,H} \geq n_{\rm crit}$. When $n_{\rm c,H} < n_{\rm crit}$, we assume photoionization equilibrium with the UV background, and that the neutral fraction of hydrogen by number is given by $x_{\rm HI} =\alpha_{\rm B}n_{\rm c,H}/\Gamma_{\rm bg}$. Here, $\alpha_{\rm B}=2.6 \times 10^{-13}(T_c/10^4)^{-0.7}$ cm$^3$ s$^{-1}$ denotes the case-B recombination coefficient and $\Gamma_{\rm bg} =5 \times 10^{-13}$ s$^{-1}$ denotes the photoionization rate \citep{FG08}.

To complete the description of our outflow model, we need to assume
$v_{\rm c}(r)$. As discussed previously, this velocity profile is not
well known, and depends sensitively on when the cold gas shells
fragment, and on whether cold clouds form from the hot wind as a
result of thermal instabilities. We follow the empirical approach of
Steidel et al. (2010), who assumed that the {\it acceleration} of the
cold clumps scales as $a_c(r) =A r^{-\alpha}$, which results in a
velocity profile of the form

\begin{equation}
v_c(r)=\Big{(}\frac{2A}{\alpha-1}\Big{)}^{0.5}(r_{\rm
  min}^{1-\alpha}-r^{1-\alpha})^{0.5},
\end{equation} for $\alpha>1$ \citep{Steidel10}, where $A$ is a constant that sets the velocity at $r \rightarrow \infty$, $v_{\infty}$. That is, $v_{\infty}=\sqrt{2Ar_{\rm min}^{1-\alpha}/(\alpha-1)}$.  
We do not consider models with $\alpha \leq 1$ in this paper (see
below)\footnote{In the simulations of \citet{DS08}, the wind velocity
  increases with $r$ simply because the gas at a given radius has a
  velocity that is close to the minimum velocity it must have had, to
  reach that radius in the finite time since the launch of the
  wind. Having $v_c(r)$ increase with $r$ does therefore not solely
  represent models in which the clumps accelerate with radius.}.

Formally, our model has thus ten parameters which include: $r_{\rm
  min}$, $\alpha$, $v_{\infty}$, $T_{\rm h,0}$, $T_{\rm c,0}$, $\gamma_{\rm h}$, $\gamma_{\rm c}$, $\eta_{\rm c}$, $\eta_{\rm
  h}$ and $N_{\rm clump}$. For most of these parameters  we have
decent constraints, and they are not free. For example,
\citet{Steidel10} inferred from their observations that $1.15 < \alpha
< 1.95$, $v_{\infty}=800$ km s$^{-1}$ and adopted $r_{\rm min}=1$ kpc,
which is close to  theoretical estimates of the blow-out radius
\citep{MB09}.  The observationally inferred hot wind temperatures lie
in the range $T_{\rm h}=10^6-10^7$ K for dwarf galaxies
\citep{Martin99}, but could be larger by a factor of $\sim 10$ in
starburst galaxies \citep[e.g.][]{SH09}. Cold gas at temperatures
$T_{\rm c} > 10^4$ K would efficiently cool down to $T_{\rm c} \sim
10^4$ K, below which gas cooling becomes less efficient. Further
cooling is possible because of metals, but it is unclear to which
temperatures the gas can cool in the cold clumps, and to what extent
the clumps are heated as a result of their interaction with the hot
wind (and/or as a result of photoheating). We will consider values of
$\log T_{\rm c}=2-4$. \citet{OD06} could reproduce the observed amount
of \ion{C}{IV} absorption in quasar spectra at z=2-5 with large scale
(momentum-driven) outflows arising from star forming galaxies, for a
{\it total} mass loading factor $\eta \equiv \eta_{\rm c} +\eta_{\rm
  h}=\sigma_0/\sigma$. Here, $\sigma$ denotes the velocity dispersion
of the galaxy, and $\sigma_0=150$ km s$^{-1}$ \citep{OD08}. The
observed dispersion of the H$\alpha$ line in the galaxies that were
used to construct the stacked Ly$\alpha$ image, is $\sigma_{{\rm
    H}\alpha}\sim 100-150$ km s$^{-1}$ (see Fig~4 of Steidel et
al. 2010). Under the reasonable assumption that $\sigma_{{\rm
    H}\alpha}$ provides a good measure of $\sigma$, we require that
the total mass loading factor is close to unity. However, we caution
that direct observational constraints on $\eta$ are uncertain by a
factor of at least a few \citep[][and references therein]{Genel}.
The adiabatic index of the cold gas is $1 \leq \gamma \leq 5/3 $,
where $\gamma=1$ [$\gamma=5/3$] corresponds to isothermal [adiabatic]
gas. 

In this paper, we only consider isothermal cold clouds,
i.e. $\gamma_{\rm c}=1$, because we found that for models with
$\gamma_{\rm c}> 1.2$ generally the cold clumps are compressed too
much, which gives rise to EW-$b$ curves that decline too steeply with
$b$. We also only consider models with either $\gamma_{\rm h}=1$ and
$\gamma_{\rm h}=1.2$. As we will show, in models with $\gamma_{\rm h}
=1.2$ the cold clumps expand more rapidly, which causes $n_{\rm c,H}<
n_{\rm crit}$, and the HI column density of the cold clumps declines
too fast for significant scattering of Ly$\alpha$ photons. Finally, we
only consider $N_{\rm clump}=10^6$ and $N_{\rm clump}=10^5$. This
choice for $N_{\rm clump}$ translates to clump masses in the range
$m_{\rm c}=10^4-10^5 M_{\odot}$ (see Table~\ref{table:models}). We
stress that our main results are insensitive to the adopted value for
$N_{\rm clump}$.

\subsection{Constraining the Free Parameters of the Model with Absorption Line Data}
\label{sec:model2}

As was discussed above, each outflow model contains ten parameters, seven of which are allowed to vary within a reasonably well known range. Each model is therefore parameterized by the 7-D parameter vector ${\bf P}$ $\equiv (r_{\rm min},\alpha,v_{\infty},T_{\rm h,0},T_{\rm c,0},\eta_{\rm c},\eta_{\rm h})$.
For each ${\bf P}$ we
obtain a distribution of cold clumps that contain neutral atomic
hydrogen. \citet{Steidel10} have measured the average Ly$\alpha$
absorption line strength at impact parameter $b$ from galaxies by
analyzing the spectra of background galaxies. We can use these
observations to constrain the components of ${\bf P}$. 

\begin{figure}
\vbox{\centerline{\epsfig{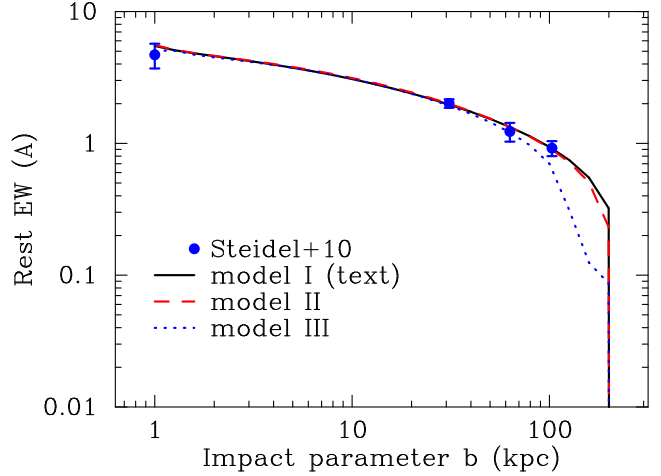}}}
\vspace{0mm}
\caption[]{This figure shows the mean absorption line strength -
  quantified by the restframe equivalent width- in the Ly$\alpha$ line
  as a function of impact parameter $b$. {\it Blue filled circles}
  indicate the observations by Steidel et al. (2010). The three lines
  indicate EW as a function of $b$ for the three models ({\bf model
    I}-{\bf III}) for the large scale outflow (see text for details on
  the model). These models are used as input to our Ly$\alpha$
  radiative transfer calculations. } 
\label{fig:ewb}
\end{figure}

Specifically, \citet{Steidel10} have measured mean absorption
equivalent width (rest frame) in the Ly$\alpha$ line at 4 impact
parameters. These measurements are shown as {\it filled blue circles}
in Figure~\ref{fig:ewb}. The equivalent width (at impact parameter
$b$) is defined as

\begin{equation}
{\rm EW}(b)=\lambda_{\alpha} \frac{\Delta
  \nu_{\alpha}}{\nu_{\alpha}}\int_{-\infty}^{+\infty} dx
(1-\exp[-\tau(x,b)]).
\label{eq:ewvsb}
\end{equation} The integral is over the dimensionless frequency $x$ (see Table~\ref{table:symbols}), and $\exp[-\tau(x,b)]$ denotes the fraction of the flux at frequency $x$ that is transmitted to the observer.

Evaluating this transmission is more complicated for a clumpy medium
than for a homogeneous medium. For example, in the hypothetical case
of $\mathcal{N}_{\rm clump}(b)=0.1$ we expect only $10\%$ of all
sightlines with impact parameter $b$ to pass through a clump, and the
outflow is transparent for the remaining $90\%$ of the sightlines. The
absorption line strength is then EW$(b)=$EW$_{\rm clump}(r)/10$, where
EW$_{\rm clump}(r)$ denotes the equivalent width as a result of
absorption by a single clump.  In the more general case, the
transmission $\exp[-\tau(x,b)]$ is a product of the transmission by
individual segments along the sightline. That is
$\exp[-\tau(x,b)]=\prod_{i=1}^{N_s}\exp[-\tau(x,b,s_i)]$. Here,
$\exp[-\tau(x,b,s_i)]$ denotes the fraction of the flux that is
transmitted by clumps in line segment `i', that lies at
line--of--sight coordinate $s_{\rm i}$, which has length $\Delta s_i$,
and which lies a distance $r_{\rm i} \equiv \sqrt{b^2+s^2_i}$ from the
galaxy. This transmission $\exp[-\tau(x,b,s_i)]=p_{\rm
  clump,i}\exp[-\tau_{\rm clump}(x')]+1-p_{\rm clump,i}$. Here,
$p_{\rm clump}$ denotes the probability that a clump lies on line
segment `i', and $\exp[-\tau_{\rm clump}(x')]$ denotes the total
fraction of the flux that is transmitted by the clump. Note that
because of the outflow velocity of the clump, $x'$ is related to $x$
via a Doppler boost (see below). The probability that line segment `i'
contains a clump is given by\footnote{Formally, we do not allow the
  clumps to overlap in our Ly$\alpha$ Monte-Carlo calculations, which
  modifies the probability $p_{\rm clump,i}=f_{\rm c}(r)\Delta s_i
  \times p({\rm no \hs clump \hs at}\hs \Delta s \leq 2R_{\rm c})$
  $\sim f_{\rm c}(r)\Delta s_i \times [1-4f_{\rm c}(r)R_c(r)]$.  We
  have explicitly checked that this difference makes no difference in
  practice because of the low volume filling factor for the cold
  clumps.} $p_{\rm clump,i}=f_{\rm c}(r)\Delta s_i$, and we obtain
\begin{eqnarray}
\exp[-\tau(x,b)]=\nonumber \\  \prod_{i=1}^{N_{\rm s}}\Big{(}\Delta s_i
f_{\rm c}(r_i) \exp[-N_{\rm c,HI}(r_i) \sigma_{\alpha}(x')]+1-\Delta
s_i f_{\rm c}(r_i)\Big{)}.
\label{eq:trans}
\end{eqnarray} Here, line segment `i' covers the range $[s_{\rm min}+(i-1)\Delta s_{\rm i},s_{\rm min}+i\Delta s_{\rm i}]$, where $s_{\rm min}=-\sqrt{r^2_{\rm max}-b^2}$ and $\Delta s_{\rm i}=2|s_{\rm min}|/N_{\rm s}$. Furthermore, $\sigma_{\alpha}(x')$ denotes the Ly$\alpha$ absorption cross-section evaluated at frequency $x'$ in the frame of the clump, i.e. $x'=x-\frac{v_c(r)}{v_{\rm th}} {\bf e}_{\rm s}\cdot {\bf e}_{\rm r}$. Here, ${\bf e}_{\rm s}$ [${\bf e}_{\rm r}$] denotes a unit vector along the line of sight toward the galaxy [along the radial vector]. We stress that the background galaxies do not provide sight{\it-lines} through the CGM of the foreground galaxy. Because of their finite physical sizes, they instead probe `sight-cylinders' of some radius $r_{\rm cyl}$ through the CGM. However, it is easy to show that Eq~\ref{eq:trans} also applies to cylinders of radius $r_{\rm cyl}$ provided that $r_{\rm cyl} \ll b$, which is the case for the observations at $b \gsim 30$ kpc since $r_{\rm cyl}$ is $\sim$ a few kpc \citep[e.g.][]{Law12}. Formally, this formalism is not correct at $b \sim 0$. However, we have explicitly verified that a more detailed calculation\footnote{In the case of a cylinder, we replace $p_{\rm clump,i}$ with the fraction of the area of the background galaxy, $A_{\rm bg}$, that is covered by clumps in cylinder segment `i', which is given by $\frac{2 \pi \Delta s_{\rm i}}{A_{\rm bg}}\int_o^{r_{\rm cyl}}dx\hs x \hs n_{\rm c}(u)A_{\rm c}(u)$, where $u^2=x^2+s_{\rm i}^2 $.} reduces the predicted EW by only $\lsim 20\%$ at $b=0$ for $r_{\rm cyl}\leq 5$ kpc.

To find models that fit the data of Steidel et al. (2010) we use a
Markov-Chain Monte-Carlo simulation to probe the parameter space
spanned by ${\bf P}$. Our exploration of the parameter space is rather
simple: our goal is to find models that provide a good fit to the
data, whether these models are physically plausible in the context of
our model, and to explore whether these same models can give rise to
the observed Ly$\alpha$ halos. We will not present probability
distribution functions for the elements of ${\bf P}$, and will not
explore the correlations that exist between them. Given the simplified
nature of the model, this would distract from the main purpose of our
analysis.

 From the Markov-chains\footnote{We generate 10 chains that contain 2000 steps
  and simply select the best--fit model from all ten chains. For each step
  we compute the likelihood $\mathcal{L}({\bf
    P})=\exp(-\chi^2/2)P_{\rm prior}({\bf P})$, where
  $\chi^2=\sum_{i=1}^{4}({\rm EW}_{\rm mod,i}-{\rm EW}_{\rm
    obs,i})^2/\sigma_{\rm EW,i}^2$. We assume flat priors on  $T_{\rm
    c}$, $\eta_{\rm h}$, $\eta_{\rm c}$, but restrict ourselves to the
  range $2 < \log T_{\rm c,0} < 4$, $0.1 < \eta_{\rm h}<10$, $0.1 <
  \eta_{\rm c}<10$ (i.e. outside this range, the prior probability is
  set to zero). We assume Gaussian priors for $r_{\rm min}$
  [$(\bar{r},\sigma)=(1.0,1.0)$], $\alpha$
  [$(\bar{\alpha},\sigma)=(1.5,0.4)$], $v_{\infty}$
  [$(\bar{v},\sigma)=(800.0,100.0)$], and $\log T_{\rm h,0}$
  [$(\bar{T},\sigma)=(7.0,0.5)$], but restrict ourselves to $0< r_{\rm
    min}< 3.0$ kpc, $0.5 < \alpha <2.0$ \citep{Steidel10}, $500 <
  v_{\infty} < 1000$ km s$^{-1}$,  and $5.0 < \log T_{\rm h,0} < 8.0$. } we select three models, which we denote with {\bf model I}-{\bf model III}.  We summarize the parameters of these models in Table~\ref{table:models}.  Figure~\ref{fig:ewb} compares the observed EW as a function of $b$ to the predictions by the models. The {\it black solid line}, {\it red dashed line}, and {\it blue dotted line} represent {\bf model I}, {\bf model II}, and {\bf model III} respectively. All models clearly provides a good fit to the data.

\begin{table*}
\caption{Outflow model parameters that we adopt in our models.}  \centering.
\begin{tabular}{l l l l l l l l l l l l}
\hline\hline
  Model & $r_{\rm min}$ (kpc) & $\alpha$ & $v_{\infty}$ (km s$^{-1}$) & $T_{\rm h,7}$ & $T_{\rm c,4}$ &  $\eta_{\rm h}$ & $\eta_{\rm c}$ & $N_{\rm clump}$ &$\gamma_c$ & $\gamma_{\rm h}$ & $m_{\rm c}$ ($M_{\odot}$)\\
  \hline
   \multicolumn{12}{|c|}{Models in which $a_c(r) = A r^{-\alpha}$ (\S~\ref{sec:cmod}).}\\
  \hline
  {\bf model I} & 1.0 & 1.4 &780 & 1.0 & 0.3 & 0.6 & 0.7 & $10^6$ & 1.0 & 1.0 & $8 \times 10^3$\\
  {\bf model II} & 1.0 & 1.5 &790 & 1.0 & 0.5 & 0.9 & 1.1 & $10^5$ & 1.0 & 1.0 & $1.2 \times 10^5$\\
  {\bf model III} & 1.2 & 1.4 &834 & 0.9 & 0.7 & 3.6 & 0.9 & $10^5$ & 1.0 & 1.2 & $9 \times 10^4$\\
\hline
   \multicolumn{12}{|c|}{Models in which $a_c(r) = A r^{-\alpha} - GM(r)/r^2$ (\S~\ref{sec:vel})}.\\
   \hline
 {\bf model IV-V} & 1.0 & 1.4 & N/A & 1.0 & 0.9 & 0.2 & 0.9 & $8.5 \times10^5$ & 1.0 & 1.0 & $1 \times 10^4$\\
   \hline\hline
\end{tabular}
\label{table:models}
\end{table*}

\begin{figure*}
\vbox{\centerline{\epsfig{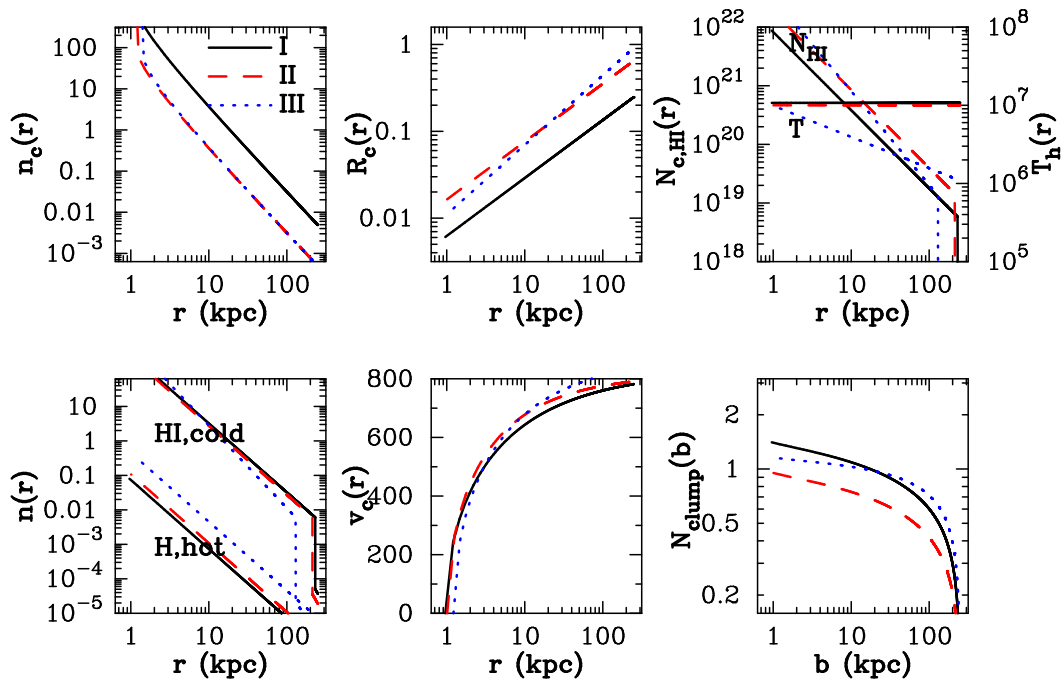}}}
\vspace{0mm}
\caption[]{The plots give a more detailed look at {\bf models I}-{\bf
    III} of the cold clumps in the large scale outflow. We show the
  number density of clumps $n_{\rm c}(r)$ in proper kpc$^{-3}$ in the
  {\it upper left panel}. The {\it upper central panel} shows the
  clump radius in proper kpc as a function of $r$. The {\it upper
    right panel} shows the HI column density (in cm$^{-2}$) through an
  individual clump located at radius $r$. The {\it lower right panel}
  shows the number density of HI atoms in the clump (in cm$^{-3}$),
  while the {\it lower central panel} shows the velocity profile (in
  km s$^{-1}$). Finally, the {\it lower right panel} shows the number
  of clumps a random sightline with impact parameter $b$ intersects.} 
\label{fig:models}
\end{figure*} 

Figure~\ref{fig:models} shows some properties of the cold clumps for
all three models, where we use the same line style and color
representation as in Figure~\ref{fig:ewb}. We discuss the clump
properties in more detail below:
\begin{itemize}

\item  {\bf Model I}: The clump radii increase from $R_{\rm c}< 0.01$
  kpc to $R_{\rm c} \sim 0.1$ kpc at r=100 kpc. The HI column density
  of the clumps falls from $N_{\rm c,HI}\sim 2 \times 10^{20}$
  cm$^{-2}$ at r=10 kpc to $N_{\rm HI} \sim 10^{19}$ cm$^{-2}$ at
  r=100 kpc. The {\it lower left panel} shows that the HI number
  density stays above $n_{\rm crit}$ all the way out to $r\sim 200$
  kpc, and that then the HI number density drops off fast. The {\it
    central lower panel} shows that the outflow velocity of the clumps
  increases rapidly at small radii, and that it barely increases
  further at r$\gsim 10$ kpc (this plot is also shown in Fig~23 of
  Steidel et al. 2010). Finally, the {\it lower right panel} shows
  that a random sightline at impact parameter $b$ intersects $\sim 1$
  clump out to $b=100$ kpc, after which it decreases rapidly. 

\item {\bf Model II}: Most differences between this model and {\bf
  model I} are easily understood: the number density of clumps is
  lower by a factor of $10$ as a result of the lower total number of
  clumps. In order to yield the same absorption line strength, the
  decrease in $n_c(r)$ is compensated for by an increase of their
  radii ({\it upper central panel}). The HI number density in the cold
  clumps is set entirely by the properties in the hot gas, and hence
  remains identical ({\it lower left panel}). As a result of the
  unchanged HI number density and the enhanced clump radii, the total
  HI column density is correspondingly larger ({\it upper right
    panel}). Finally, because of the enhanced HI column density of
  individual clumps, $\mathcal{N}_{\rm clump}(b)$ must be smaller for
  {\bf model II} in order to reproduce the observed EW$(b)$ ({\it
    lower right panel}).

\item {\bf Model III}: The number density of clumps is the almost
  identical to that of {\bf model II}, which is because this number
  density is determined mostly by the total mass outflow rate of cold
  gas (comparable for both models) as well as the velocity profile
  (also comparable for both models). The temperature of the hot gas
  decreases with radius from $T_{\rm h} \sim 10^7$ K to $T_{\rm h} \sim 10^6$ K in this model ({\it upper right panel}), pressure equilibrium also
  implies that the number density of HI atoms decreases faster in the
  model ({\it lower left panel}).  Because we fixed the clump mass,
  the clump radius is therefore increasing more rapidly with radius in
  {\bf model III} than in {\bf model II} ({\it upper central panel}). 

\end{itemize}

\subsection{Generating a Random Realization of Clumps}
\label{sec:clump3}
\begin{figure}
\vbox{\centerline{\epsfig{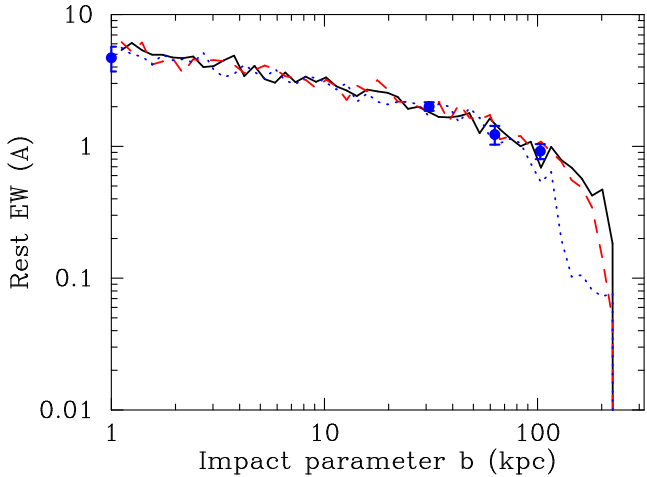}}}
\vspace{0mm}
\caption[]{Same as Fig~\ref{fig:ewb}, but for random realizations of
  {\bf model I}--{\bf model III}. This Figure illustrates that we have
  generated representative random realizations of our best-fit
  models. It also provides an independent check of our predicted EW as
  a function of $b$ in Eq~\ref{eq:trans}.} 
\label{fig:ewbran}
\end{figure} 
We generate random realizations of {\bf model I}--{\bf model III} as
follows. For each of the $N_{\rm clump}$ clumps in our model, we first
generate a random unit vector ${\bf e}_{\rm i}$. Then we generate a
random radial coordinate $r_{\rm i}$ of the clump from

\begin{equation}
\mathcal{R} = \int_0^{r_{\rm i}} dr\hs 4 \pi r^2 n_{\rm c}(r),
\label{eq:nothing}
\end{equation} where $\mathcal{R}$ is a random number between 0 and 1. The center of the clump `i' is then given by ${\bf x}_{\rm i} \equiv r_{\rm i}{\bf e}_{\rm i}$. We then check whether clump `i' overlaps with any of the previously generated i-1 clumps. In case it does, we generate a new random unit vector ${\bf e}_{\rm i}$, until clump `i' does not overlap with any of the other clumps. We then proceed to generating the position of clump `i+1'. The velocity, temperature, radius, HI number density, and dust content are specified fully for a given $r_{\rm i}$ and a given model (the dust content will be discussed in \S~\ref{sec:dust}).

Once we have random realizations for each model, we shoot random
sightlines at a range of impact parameters (100 sightlines at each
impact parameter) and compute the mean Ly$\alpha$ absorption strength
as a function of impact parameter. We show the results of this
exercise in Figure~\ref{fig:ewbran}, where we use the same line colors
and style to represent the different models as in the previous
plots. Our discrete realizations of clumps also give rise to
Ly$\alpha$ absorption at levels that are in excellent agreement with
the data. This provides an independent check of the accuracy of
Eq~\ref{eq:trans} and that we have generated representative random
realizations of our best-fit models.

\section {The Ly$\alpha$ Transfer Model}
\label{sec:lyart}

\subsection{Monte-Carlo Calculations}
\label{sec:lyamc}

Our Monte-Carlo Ly$\alpha$ radiative transfer code was developed,
tested and described in more detail in \citet{D06}. For our current
work we modify the code in several ways:

\begin{itemize}

\item We can follow the transfer through an arbitrary distribution of
  spherical clumps. Each clump is assigned a location, a radius,  a
  hydrogen number density, a dust number density, a temperature, and a
  (radial) velocity. This differs from the applications presented in
  \citet{D06}, where we focused on single spherically symmetric gas
  clouds. In \citet{D06} the single clump was divided into a large
  number of spherically concentric shells, to each of which we
  assigned a velocity, HI density, and temperature. In this work, we
  do not allow for gradients of temperature etc. across the clump
  (although our code can be modified for this purpose).

\item We include the impact of dust on the radiative transfer process
  following Laursen et al. (2009). We assume an `SMC' type frequency
  dependence of the dust absorption cross-section. However, this
  frequency dependence is practically irrelevant over the narrow range
  of frequencies which are covered by Ly$\alpha$ photons (see Laursen
  et al. 2009 for details). When Ly$\alpha$ photons scatter off a dust
  grain, we assume that the scattering is described by a
  Henyey-Greenstein phase function with asymmetry parameter $g=0.73$
  \citep[][and references therein]{Laursen09}. Scattering of UV
  radiation by SMC type dust gives rise to little linear polarization
  \citep[see Fig~5 of][]{Draine03}, and for simplicity we shall assume
  that scattering by dust grains does not polarize Ly$\alpha$
  radiation. For the work presented in this paper we will assume that
  the albedo, which denotes the ratio of the scattering to the total
  cross-section, is $A=0.0$ (see \S~\ref{sec:dust}). We assume that
  the relation between dust {\it absorption} opacity $\tau_{\rm D,a}$
  and the color excess $E_{B-V}$ is given by $\tau_{\rm
    D,a}=10.0E_{B-V}$ \citep[also see][]{Verhamme06}. 

\item Because we want the option to study clumpy outflows that are not
  spherically symmetric, we generate surface brightness profiles with
  the so-called `peeling algorithm', or the `next event
  estimator'. This technique has been employed in many previous
  studies of Ly$\alpha$ transfer
  \citep[e.g.][]{Zheng02,Ca05,Ta06,Laursen07,FG10,K10,Barnes11}. A
  more detailed description of how we generate images can be found in
  Appendix~\ref{app:nee}.

\item As in \citet{DL08} we compute polarization of scattered
  radiation. Polarization calculations have thus far focused solely on
  spherically symmetric gas distributions.

\end{itemize}

We point out that our code allows us to perfectly `resolve' the
Ly$\alpha$ transfer inside of clumps with arbitrary small radii (in this case $R_c \lsim 10$ pc)
(see \S~\ref{sec:model2}) within  our large (diameter $\sim 500$ kpc)
outflow.

\subsection{Analytic Calculations}
\label{sec:ana}

Under the assumption that the Ly$\alpha$ photons scatter only
once--which is reasonable as we will argue below-- we can compute the
surface brightness profile [$S(b)$] as well as the polarization
profile [$\mathcal{P}(b)$] analytically as
\begin{eqnarray}
S(b)=S_l(b)+S_r(b)
\\ \mathcal{P}(b)=\frac{|S_l(b)-S_r(b)|}{S_l(b)+S_r(b)}
\end{eqnarray}, where $S_l(b)$ and $S_r(b)$ denote polarized fluxes \citep{RL99,DL08}, which are given by

\begin{eqnarray}
\left.\begin{array}{ll} S_l(b)\\ S_r(b)
     \end{array}
\right\} =\Big{(}\frac{{\rm kpc}}{{\rm asec}}
\Big{)}^2\int_{-\infty}^{\infty}dx\int_{-\infty}^{\infty}
ds\hs\frac{r}{b} \times \frac{3}{4} \times \\ \nonumber s_{\rm
  in}(x,b,s)\times f_{\rm c}(s,b)\times \big{(}1- \exp[-\tau_{\rm
    clump}(x',s,b)]\big{)}\times \\ \nonumber \times f_{\rm esc}(x'',b,s)
\times  \left\{\begin{array}{ll} \mu^2\\ 1\nonumber
     \end{array},
     \label{eq:sb}
\right.
\end{eqnarray} where the term (kpc/asec)$^2$ converts the surface brightness into units erg s$^{-1}$ cm$^{-2}$ arcsec$^{-2}$. Furthermore, $s_{\rm in}(x,b,s)$ denotes the total incoming flux at location $(b,s)$ and frequency $x$, where $s$ denotes the line--of--sight coordinate (see Fig~\ref{fig:b}).  We can write $s_{\rm in}(x,b,s)=s_0 \frac{n(x)}{4 \pi r^2}\exp(-\tau[s,b,x])$, in which $s_0$ denotes the total observed Ly$\alpha$ flux of the source if no scattering occurred at all\footnote{The flux $s_0$ relates to the intrinsic luminosity, $L_{\alpha}$, of the source simple as $s_0 \equiv L_{\alpha} 4 \pi d^2_{\rm L}(z)$, where $d_{\rm L}(z)$ denotes the luminosity distance to redshift $z$.}, $r \equiv \sqrt{s^2+b^2}$, and $\exp(-\tau[s,b,x])$ denotes the total fraction of the flux at frequency $x$ that has {\it not} been scattered out of the line of sight yet. The term $\exp(-\tau[s,b,x])$ is computed as in Eq~\ref{eq:trans}. The optical depth $\tau_{\rm clump}(x',s,b)=N_{\rm HI,c}(r)\sigma_{\alpha}(x')$, denotes the optical depth through the clump at radius $r$ at frequency $x$. In the frame of the clump, photons of frequency $x$ appear at $x'=x-v_{\rm c}(r)/v_{\rm th}$. The last line contains the scattering phase function, in which $\mu=-s/r$, and accounts for the fact that photons are not scattered in all directions with equal probability. Finally, $f_{\rm esc}(x'',b,s)$ is the probability that photons that are scattered towards the observer, are detected. This probability can again be computed as in Eq~\ref{eq:trans}, but note that after scattering the photon's frequency has been changed to $x''=x+(\mu-1)v_c(r)/v_{\rm th}$. We present a complete derivation of this equation in Appendix~\ref{app:an}. 

\section{Results}
\label{sec:results}
In our Monte-Carlo calculations we emit $N_{\gamma}=10^5$ Ly$\alpha$
photons in the center of the clumpy outflow for each model, and
randomly draw the initial frequency of each emitted photon from a
Gaussian with a standard deviation of $\sigma=150$ km s$^{-1}$. This
is close to the typical dispersion of the observed H$\alpha$ lines in
the sample, which is about $\sigma_{{\rm H}\alpha}\sim 100$ km
s$^{-1}$ (see Fig~4 of Steidel et al. 2010). We assume that the
luminosity of the central source is $L_{\alpha}=3.4 \times 10^{43}$
erg s$^{-1}$. This luminosity corresponds to the intrinsic Ly$\alpha$
luminosity of a galaxy that is forming stars at a rate SFR=34
$M_{\odot}$ yr$^{-1}$, which corresponds to the median UV-based
dust-corrected SFR of all 92 continuum selected galaxies that were
used to create the stacked Ly$\alpha$ image.  We thus implicitly
assume that the escape fraction of Ly$\alpha$ photons from the dusty
interstellar medium into the large scale outflow is\footnote{The
  conversion $L_{\alpha}=10^{42} \times {\rm [SFR}/(M_{\odot}/{\rm
      yr})]$ applies for a Salpeter stellar initial mass function
  (IMF) and solar metallicities.  We expect a larger Ly$\alpha$
  luminosity at fixed SFR at lower gas metallicities \citep{S03}. For
  a Chabrier IMF -- and again solar metallicity -- we expect $\sim
  1.8$ times more Ly$\alpha$ luminosity at a given SFR
  \citep{Steidel10}. It may therefore be possible to get the same
  Ly$\alpha$ luminosity from the central source if the escape fraction
  is $f_{\rm esc}\sim 50\%$, which is still large.} $f_{\rm
  esc}=100$\%. The predicted surface brightness scales linearly with
$f_{\rm esc}$. The luminosity $L_{\alpha}=3.4 \times 10^{43}$ erg
s$^{-1}$ which at $z=2.65$ translates to $s_0=5.9 \times 10^{-16}$ erg
s$^{-1}$ cm$^{-2}$, which is relevant for our analytic calculations.

Throughout, we represent the observed Ly$\alpha$ surface brightness
profile by the function $S(b) =S_0 \exp [-b/b_{\rm c}]$, where $b$
denotes the impact parameter in kpc. Here, $S_0=2.4 \times 10^{-18}$
erg s$^{-1}$ cm$^{-2}$ arcsec$^{-2}$, and $b_{\rm c}=25.2$ kpc. This
function provides an accurate fit to the average Ly$\alpha$ observed
in the full sample of 92 continuum-selected galaxies with Ly$\alpha$
imaging \citep{Steidel11}, and is shown as {\it red solid lines} in
the Figures below.

We represent results from our Monte-Carlo calculations with {\it black
  filled solid circles}, which contain errorbars. We obtain these
points by averaging the six surface brightness (and polarization)
profiles that we obtain by viewing the clump distribution from six
different directions (see \S~\ref{sec:lyamc}). Uncertainties on the
these points indicate the standard deviation from this average. 

\subsection{Dust Free Clumpy Outflows}
\label{sec:dustfree}

\begin{figure*}
\vbox{\centerline{\epsfig{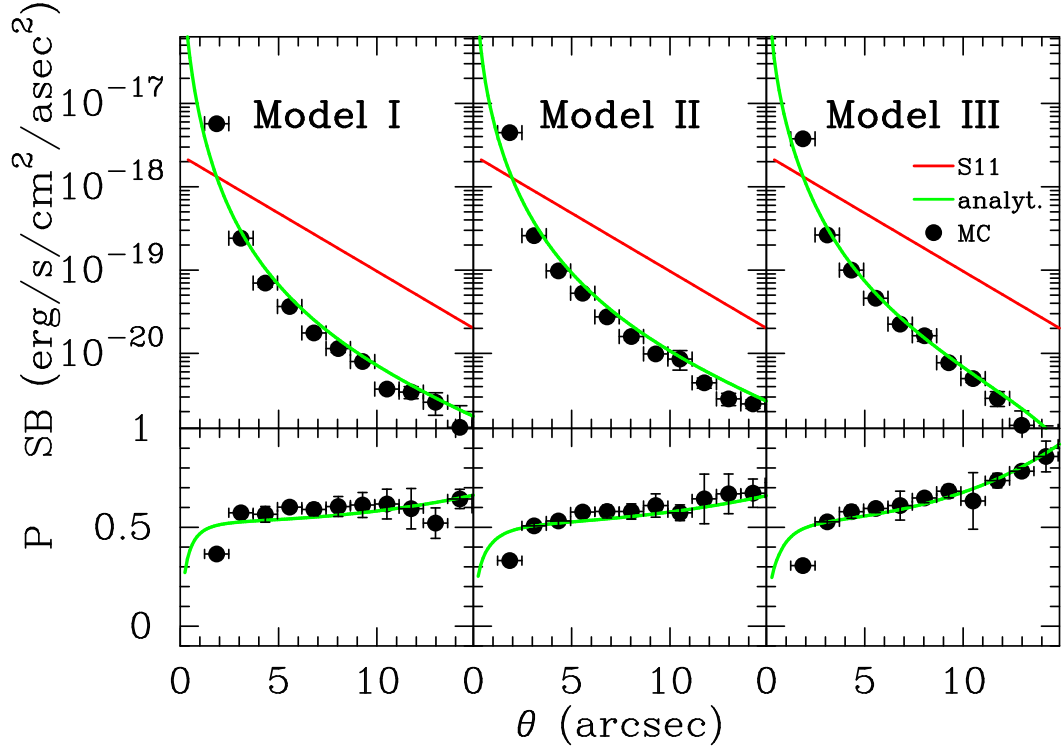}}}
\vspace{0mm}
\caption[]{This Figure shows a comparison between the `predicted' and
  observed Ly$\alpha$ line profiles for our three models. The {\it red
    solid lines} show the surface brightness profile that was observed
  by Steidel et al. (2011). The {\it solid green lines} show the
  surface brightness profiles as given by Eq~12 which assumes
  that photons scatter only once. The {\it filled black circles} show
  the results from our Monte-Carlo calculations. This Figure shows
  that the scattered Ly$\alpha$ radiation in our models gives rise to
  Ly$\alpha$ halos that are more compact (i.e. centrally concentrated)
  and fainter by $\sim$ an order of magnitude at $\theta \gsim 5$
  arcsec. The {\it lower panel} shows the linear polarization as a
  function of impact parameter. We find that scattering through clumpy
  outflows can give rise to high levels of polarization. This figure
  also shows that there is very good agreement between our analytic
  and Monte-Carlo calculations, which is a consequence of the fact
  that photons typically scatter in only one clump (or none at all).} 
\label{fig:sb1}
\end{figure*}

 The {\it top panels} of Figure~\ref{fig:sb1} compare the observed
 Ly$\alpha$ surface brightness profile  (in erg s$^{-1}$ cm$^{-2}$
 arcsec$^{-2}$) of \citet{Steidel11} with the `predicted' surface
 brightness profiles obtained from the Monte-Carlo ({\it filled black
   circles}) and analytic ({\it green solid lines}) for {\bf model
   I}-{\bf model III}. These figures show clearly that none of the
 models can reproduce the observations:  all three models have too
 much flux coming from $\theta < 2 $ arcsec, although the actual
 observations also show a significant excess over the exponential
 fitting function at these impact parameters. This difference can be
 reduced by including dust (see \S~\ref{sec:dust}). The most important
 difference however, is at large impact parameters ($\theta \gsim 5$
 arcsec, or $b \gsim 40$ kpc), where {\it our predicted surface
   brightness profiles are low by an order of magnitude}.
 
The fact that our model surface brightness profiles are so much
fainter is easy to understand: the {\it central lower panel} of
Figure~\ref{fig:models} shows that at $r > 10$ kpc, the cold clouds
are moving away from the central Ly$\alpha$ source at $v_c \gsim 600$
km s$^{-1}$.  The majority (95\%) of Ly$\alpha$ photons will therefore
enter the clumps with an apparent redshift of $\Delta v = 600 \pm
2\sigma=300-900$ km s$^{-1}$. In order for the clouds to be opaque
($\tau>1$) to Ly$\alpha$ photons, we must have $N_{\rm HI}\sim 3
\times 10^{19}-3 \times 10^{20}$ cm$^{-2}$. For most of our models,
this requirement is met. However, the observed Ly$\alpha$ absorption
line strength at $b=30$ kpc (see e.g. Fig~\ref{fig:ewb}) is EW$_{\rm
  obs}(b=30\hs{\rm kpc})\sim 2$ \AA, which is smaller than than the
absorption equivalent that would be produced by a single clump with
this HI column density. The number of such clumps along the line of
sight is therefore small (order unity, see the {\it lower right panel}
of Fig~\ref{fig:models}), and because the radii of the clumps $R_c \ll
r$, {\it the majority of photons escape from the circum galactic
  medium without encountering the outflowing, cold clumps.}

This last point is also illustrated nicely by the fraction of photons
that never scatter in the outflow, denoted by $f_{\rm ns}$: in {\bf
  model I} we find that $f_{\rm ns} \sim 0.75$. That is, the majority
of photons do not scatter off the cold clumps that give rise to the
observed absorption. For {\bf model II}, we have $f_{\rm ns}\sim
0.80$, while for {\bf model III} we have $f_{\rm ns}=0.43$. An
immediate implication of this finding is that the photons that do not
scatter should be observed as a Ly$\alpha$ point source of equal or
larger luminosity than that of the Ly$\alpha$ halo. In the
observations, the luminosity of the Ly$\alpha$ halo exceeds that of
the central source by about a factor of $\sim 5$, also in disagreement
with our model.

Those photons that do scatter in the outflow, get Doppler boosted to
lower frequencies after they escape from the clump. The probability
that the photon scatters in a second clump is then reduced further,
because they appear even further from line resonance in the frame of
the other clumps. Indeed, we find in the Monte-Carlo simulations that
photons generally scatter only in one clump, and this is the reason
why our analytic solutions for the surface brightness profiles closely
matches the ones we obtained with the Monte-Carlo method\footnote{We
  stress that we do not expect perfect agreement, mostly because:
  ({\it i}) in our dust-free Monte-Carlo calculations the total flux
  of Ly$\alpha$ photons through each radial shell is conserved (but
  redistributed along the frequency axis), while this is not the case
  for the analytic calculations; ({\it ii}) the analytic formulation
  does not properly account for radiative transfer effects inside the
  clump where the photon scatters; ({\it iii}) we construct images
  that contain pixels that are $5\times 5$ kpc$^2$ on a side, and
  binning is involved when creating azimuthally averaged surface
  brightness profiles.}.
The {\it lower panels} shows that the scattered Ly$\alpha$ radiation
is highly polarized, with the linear polarization $\mathcal{P}\gsim
40\%$ at $S \lsim 10^{-18}$ erg s$^{-1}$ cm$^{-2}$ arcsec$^{-2}$. This
high level of polarization is another consequence of the fact that the
photons typically scatter only once\footnote{The HI column density of
  clumps $N_{\rm HI} \gsim 10^{21}$ cm$^{-2}$ at $r \lsim 10$
  kpc. Photons typically scatter several times on the surface of these
  clumps before escaping from them. This suppressed our the
  polarization signal that we predict with the Monte-Carlo
  calculations at small impact parameters. We do not completely `wash
  out' the polarization signature because polarization measures the
  anisotropy in the local Ly$\alpha$ radiation field weighted by the
  photons' escape probabilities. This can lead to significant levels
  of polarization despite the fact that photons can scatter multiple
  times \citep[also see][]{DL08}.}.
\begin{figure*}
\vbox{\centerline{\epsfig{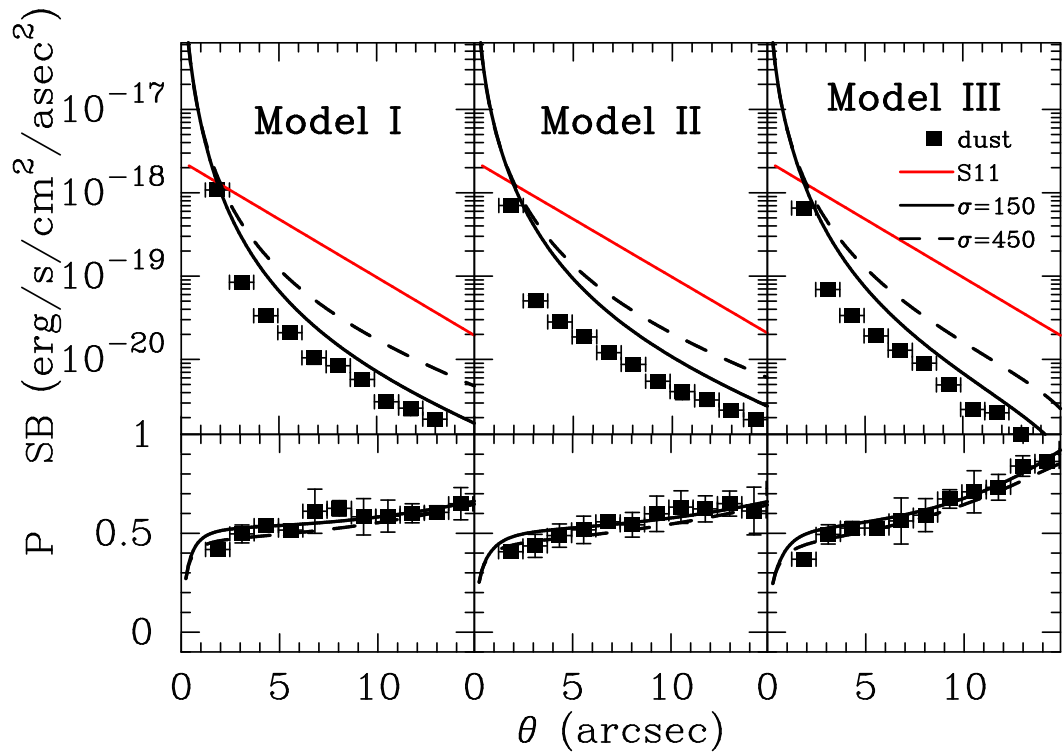}}}
\vspace{0mm}
\caption[]{This Figure shows the same as Figure~\ref{fig:sb1}. Here,
  we include models in which we assumed the initial Ly$\alpha$ line
  profile to be broader by a factor of $3$, i.e. $\sigma=450$ km
  s$^{-1}$. The resulting surface brightness profiles and polarization
  profiles are shown as the {\it black dashed lines}. We have shown
  results from our analytic calculations only. These plots show that
  significant (and probably unrealistic, see text) broadening the
  initial line profile increases the surface brightness level of the
  scattered Ly$\alpha$ halos, but not nearly enough to account for the
  observed Ly$\alpha$ halo profiles. The {\it filled black squares}
  show the result of models in which we include (a generous amount of)
  dust. Dust affects the inner regions of the surface brightness
  profiles most strongly.} 
\label{fig:sb2}
\end{figure*} 

Figure~\ref{fig:sb2} shows how our predicted surface brightness
profiles change when we increase the width of the Ly$\alpha$ spectral
line of the central source by a factor of three to $\sigma=450$ km
s$^{-1}$. Increasing the line width enhances the fraction of photons that enter the outflow with a significant blueshift. These photons appear closer to the line resonance in the frame of the outflowing gas, and are therefore more likely scattered.
These plots show that broadening the initial line profile
increases the surface brightness level of the scattered Ly$\alpha$
halos, but not nearly enough to account for the observed Ly$\alpha$
halo profiles. Choosing even broader lines would clearly increase the
surface brightness of the Ly$\alpha$ halos more, but this choice would
probably be unphysical. Scattering of Ly$\alpha$ photons through large
columns of interstellar HI gas ($N_{\rm HI} \gsim 10^{21}$ cm$^{-2}$)
could in theory easily broaden the line even more than this, but the
line broadening as a result of scattering is limited by dust in the
ISM (and it's distribution, see Figure~8 of Laursen et
al. 2009). Moreover, \ion{H}{I} column densities of this magnitude would efficiently trap Ly$\alpha$ radiation. The radiation pressure exerted by this trapped radiation can cause the HI gas to expand outwards \citep{DL08}.
Scattering of Ly$\alpha$ photons by
this outflowing gas would broaden, but also redshift the line
\citep[e.g.][]{Ahn02,Verhamme06}. Especially for \ion{H}{I} column density of $N_{\rm HI} \sim 10^{21}$ cm$^{-2}$, this redshift can be large even for outflow velocities of only a few tens of km s$^{-1}$. 
These redshifted Ly$\alpha$ photons
would appear further from resonance in the frame of the cold clumps,
and they would less likely be scattered.  
We therefore think that our choice $\sigma=450$ km s$^{-1}$ corresponds to a reasonable upper
limit on the amount of line broadening that occurs in the interstellar
medium. Finally, the {\it filled black squares} show that the
predicted surface profiles are affected most strongly by dust in the
central regions  (see \S~\ref{sec:dust}).

The fundamental reason why we fail to reproduce the observed
Ly$\alpha$ halos is that the clumps at large radii are moving away
from the Ly$\alpha$ source too fast, thus requiring prohibitively
large HI column densities in order to remain opaque to the Ly$\alpha$
photons.  In \S~\ref{sec:vel} we explore a class of models for which
the cold clumps decelerate after some radius, as expected naturally in
models of momentum driven winds.

\subsection{Dusty Clumpy Outflows}
\label{sec:dust}

In all models, the central clumps have the largest column
densities. If the dust opacity of the clumps scales with their HI
column density, then we can suppress the observed flux from the
central regions. We can therefore `flatten' the predicted surface
brightness profile by adding dust to the clumps. We investigate this
in more detail here. We rerun {\bf model I-III} but add dust to
clumps. The total amount of dust in a clump is normalized such that
average dust absorption optical depth through a clump is $\tau_{\rm
  dust,abs} \equiv k_{\rm dust }\times (N_{\rm HI}/10^{20}\hs{\rm
  cm}^{-2})$. That is, for a column density of $N_{\rm HI}=10^{20}$
cm$^{-2}$, the total dust absorption opacity is $\tau_{\rm d}=k_{\rm
  dust}$. We will explore the impact of dust for $k_{\rm dust}=1$. For
comparison, the diffuse HI phase of the Milky-Way has $k_{\rm
  dust}=0.1$ \citep[][and references therein]{Ho06}. We choose the
larger value $k_{\rm dust}=1$ to clearly illustrate the potential
impact of dust. To maximize the impact of dust, we also assumed that
the dust grains do not scatter Ly$\alpha$ photons (i.e. $A=0$).

The {\it black filled squares} in Figure~\ref{fig:sb2} show the
predicted surface brightness and polarization profiles in the presence
of dust. We indeed find that the surface brightness profiles are
suppressed most at small impact parameters. Also, the predicted
polarization is affected very little. Note that for non-zero dust
scattering albedo, the radiation that was scattered by dust would be
unpolarized, which would lower the overall polarization. However, if
the Ly$\alpha$ halos were a result of scattering by dust grains, then
we would expect the UV-continuum to follow the same surface brightness
profile, which is not observed \citep{Steidel11}.

\section{Ly$\alpha$ Halos for More `Realistic'  Models}
\begin{figure}
\vbox{\centerline{\epsfig{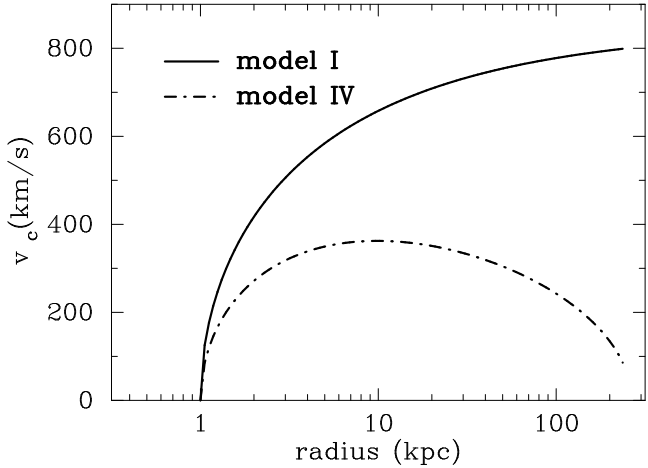}}}
\vspace{0mm}
\caption[]{This plot shows the velocity profiles that we adopt for
  {\bf model I} ({\it solid line}), and {\bf model IV} ({\it
    dot-dashed line}). In contrast to {\bf model I}, in {\bf model IV}
  gravity causes the clumps to decelerate beyond some critical radius
  $r_{\rm crit}$. As a result, clumps flow out at lower velocities in
  {\bf model IV}, which makes them more opaque to Ly$\alpha$ photons,
  which results in brighter Ly$\alpha$ halos (text).} 
\label{fig:vel}
\end{figure}

\subsection{Model IV: Adopting a More Realistic Velocity Profile}
\label{sec:vel}
\begin{figure*}
\vbox{\centerline{\epsfig{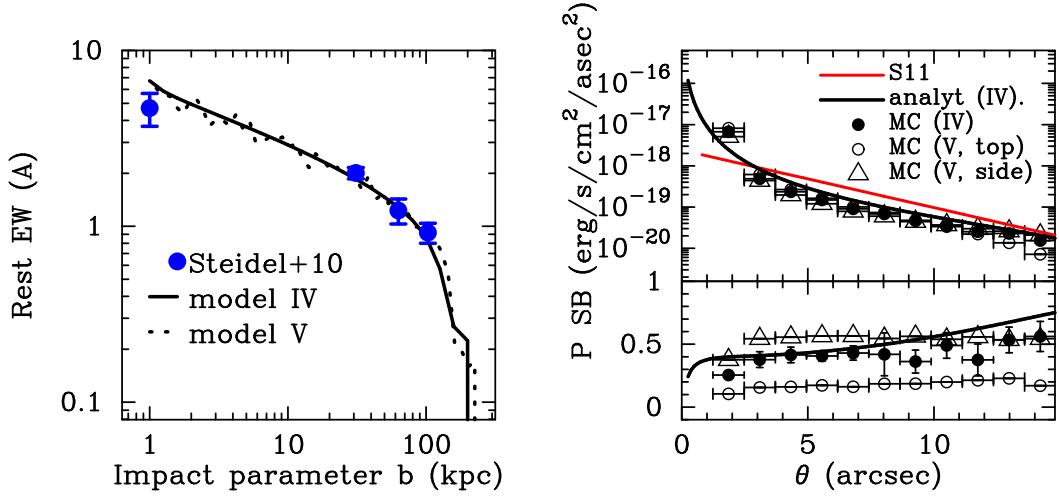}}}
\vspace{0mm}
\caption[]{The {\it left panel} shows the predicted absorption line
  strength as a function of impact parameter for {\bf model IV}, while
  the {\it right panel} shows the predicted Ly$\alpha$ surface
  brightness profiles as obtained from the Monte-Carlo simulation
  ({\it black filled circles}), and analytic calculation ({\it black
    dashed line}). This model -- in which gravity decelerates the cold
  clumps after their initial acceleration -- clearly predicts a
  surface brightness profile that is much closer to the observations
  than {\bf model I-III}.} 
\label{fig:sb4}
\end{figure*} 

In our previous models, the cold clouds accelerate as they break out
of the interstellar medium. The acceleration decreased with radius as
$a_c(r) \propto r^{-\alpha}$, where $\alpha=1.4-1.5$ in {\bf model
  I-III}. This continuous acceleration represents a `momentum--driven'
wind scenario in which the cloud's acceleration is driven by for
example ram pressure of the hot wind or radiation pressure. However,
galaxies populate the centers of gravitational wells, and the cold
clumps are subject to a gravitational force which decreases as
$\propto r^{-1}$ in the potential of an isothermal sphere. The
deceleration of clumps as a result of gravity therefore decreases
slower with radius than $a_c(r)$, and gravity dominates beyond some
some 'transition' radius $r_{\rm trans}$. This deceleration following
acceleration occurs for any model in which $\alpha > 1$. This
deceleration potentially has important implications for the predicted
surface brightness profiles of the Ly$\alpha$ halos: one of the main
reasons {\bf model I-III} significantly underpredict the Ly$\alpha$
surface brightness is that the clumps were receding from the
Ly$\alpha$ source too rapidly (see \S~\ref{sec:dustfree}).

 The goal of this section is to investigate whether we can reproduce
 the absorption line and Ly$\alpha$ halo data better in simplified
 models which take this gravitational deceleration of the cold clumps
 into account. Following Murray et al. (2005, also see Martin 2005) we
 write the momentum equation for a cold, optically thick, clump as
\begin{equation}
\frac{dv_c}{dt}=-\frac{GM(r)}{r^2}+Ar^{-\alpha},
\label{eq:de}
\end{equation} where we used the same function to describe the cloud acceleration as before (\S~\ref{sec:cmod}). The case $\alpha=2$, $A=2 \sigma^2 R_{\rm g}$ corresponds to Eq~24 of \citet{Murray05}. Here, $R_g \equiv r_{\rm min} (L/L_{\rm edd})$, where the ratio $L/L_{\rm edd}$ denotes the total luminosity of the source normalized to the Eddington luminosity of the galaxy (see Murray et al. 2005 for details). Following \citet{Murray05} we take for the model of the gravitational potential that of an isothermal sphere for which $M(r)=2 \sigma^2r/G$, in which $\sigma$ denotes the velocity dispersion. 

The solution to Eq~\ref{eq:de} is given by
\begin{equation}
v_c(r)=2\sigma\sqrt{{\rm ln}\Big{(}\frac{r_{\rm
      min}}{r}\Big{)}+\frac{A}{2\sigma^2(1-\alpha)}\Big{(}r^{1-\alpha}-r_{\rm
    min}^{1-\alpha} \Big{)}},
\label{eq:vel}
\end{equation} where we assumed the boundary condition $v_c(r_{\rm min})=0$.

In theory, it is straightforward to repeat the analysis of
\S~\ref{sec:cmod}, and do a MCMC simulation to simultaneously
constrain all model parameters including $r_{\rm min}$, $A$, and
$\sigma$ by finding the best-fit model to the absorption line data. We
have instead fixed the values $\alpha=1.4$, $r_{\rm min}=1$ kpc, which
we found to provide good fits for {\bf model I}--{\bf model III}.  We
also assumed $\sigma=150$ km s$^{-1}$, which is the value that we
assumed previously (see \S~\ref{sec:results}), and $R_{\rm g}=2.5$
(i.e. $A=5\sigma^2$). This latter choice is entirely empirical: The
{\it dot-dashed line} in Figure~\ref{fig:vel} shows that the resulting
velocity profile (Eq~\ref{eq:vel}) reaches a maximum of $v_{\rm
  c,max}\sim 350$ km s$^{-1}$ at $r\sim 10$ kpc and then decreases to
$v_{\rm c}\sim 100$ km s$^{-1}$ at $r=250$ kpc. For smaller values of
$R_{\rm g}$, the clumps would turn around and fall back onto the
galaxy, while larger values of $R_{\rm g}$ would result in negligible
deceleration. The {\it solid line} shows the velocity profile that was
adopted in {\bf model I}.
\begin{figure*}
\vbox{\centerline{\epsfig{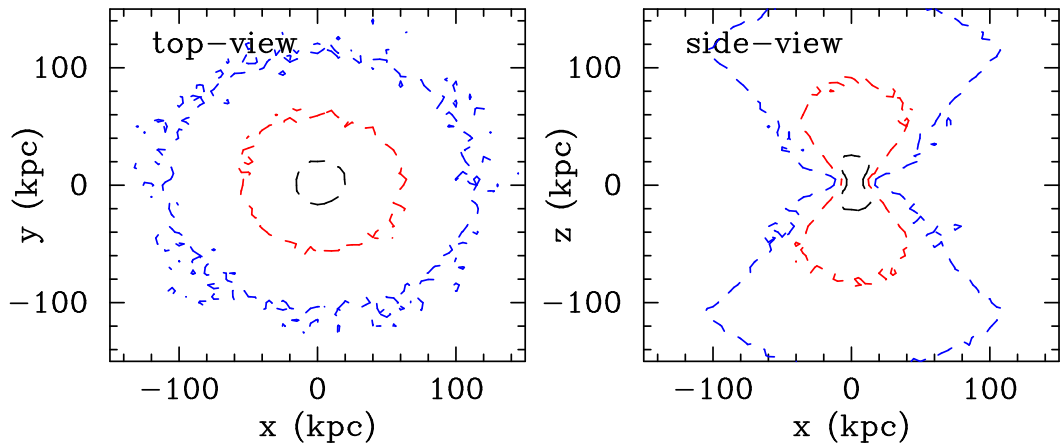}}}
\vspace{0mm}
\caption[]{This Figure shows surface brightness contours for a
  biconical outflow model as seen along the axis of the cone ({\it
    left panel}), an perpendicular to the cone axis ({\it right
    panel}). The {\it black}/{\it red}/{\it blue} contour denote a
  surface brightness level of $\log [{\rm SB}$/(erg s$^{-1}$ cm$^{-2}$
    arcsec$^{-2}$)]=-18.0/-19.0/-20.0. The opening angle of the cone
  is 90$^{\circ}$. This plot illustrates that a case in which our code works on
  non-spherical clump distributions. } 
\label{fig:contour}
\end{figure*} 

The MCMC finds the best-fit model for the parameters shown in Table~\ref{table:models}.  This model -- which
we refer to as {\bf model IV} -- is compared to the absorption line
data in the {\it left panel} of Figure~\ref{fig:sb4}. The {\it right
  panel} of Figure~\ref{fig:sb4} shows the predicted Ly$\alpha$
surface brightness and polarization profiles. It is clear that the
predicted surface brightness is significantly higher than that
predicted by {\bf model I}--{\bf model III} and agrees with the data
to within a factor of $\sim 2-3$. The enhanced surface brightness
profile is a direct result of the lower outflow velocity of the clumps
which makes them more opaque to Ly$\alpha$ photons.  The photons still
most often scatter off 1 clump, and the analytic calculation still
compares quite well to the result we obtained from the Monte-Carlo
simulation. The predicted polarization is also high for this model:
with the linear polarization reaching $P\sim 40\%$ at a surface
brightness level of SB$\sim 10^{-18}$ erg s$^{-1}$ cm$^{-2}$
arcsec$^{-2}$.

Importantly, this model still predicts that $f_{\rm ns}\sim 60\%$ of
the photons does not scatter at all in the outflow. This model
therefore also predicts that the Ly$\alpha$ halo is accompanied by a
point source of comparable luminosity, which is not observed.

\subsection{Model V: Concentrating the Outflow into Cones}
\label{sec:bipolar}

The final model that we consider -- which we refer to as {\bf model V}
-- is a biconical outflow model, in which the hot and cold gas escape
from the galaxy along two cones whose axes are parallel. The reason
that we study such a model is the observed azimuthal dependence of the
\ion{Mg}{II} at $b<50$ kpc of inclined disk-dominated galaxies at
$z=0.5-0.9$, which indicates the presence of a biconical outflows that
are aligned along the disk rotation axis \citep{Bordoloi11}. We would
like to see how our results change if we drop the assumption of having
symmetric outflows. 

We take the parameters from {\bf model IV} but now compress the
outflowing clumps into two cones. For simplicity we align the cone
axes with the $z$-axis, and assume that the opening angle is $\theta
=45^{\circ}$ (i.e. the edge of the cone intersect the $z$-axis at
$45^{\circ}$). This compression of the outflow has two implications:
({\it i}) the number density of cold clumps is enhanced by a factor of
$1/[1-\cos \theta]\sim 3.5$ at each radius, and ({\it ii}) the
pressure of the hot wind also increases by a factor of $3.5$, which
compresses the radius of the cold clouds by a factor of
$3.5^{1/3}=1.5$. The {\it dotted line} in the {\it left panel} of
Figure~\ref{fig:sb4} shows the predicted EW vs $b$ curve for a random
realization of {\bf model V}. This model also nicely reproduces the
observations without any further tuning. We have not performed an
analytic calculation of EW as a function of $b$ (as in
Eq~\ref{eq:ewvsb}), as this would require averaging over random cone
orientations.

Figure~\ref{fig:contour} shows  surface brightness contours for {\bf
  model V} when we view the outflow along the cone axis ({\it left
  panel}), and perpendicular to the cone axis ({\it right panel}). The
{\it black}/{\it red}/{\it blue contour} indicates a surface
brightness level of $\log$[SB/(erg s$^{-1}$ cm$^{-2}$ arcsec$^{-2}$)]=
-18.0/-19.0/-20.0. This Figure illustrates that our code works on
non-spherical clump distributions. For this calculation, we assume
that all Ly$\alpha$ photons escapes from the galaxy into the cones {\bf (i.e. the central source does not emit Ly$\alpha$ photons isotropically, but instead into two cones with opening angles if $45^{\circ}$)},
which is motivated by the physical picture in which the outflow has
`cleared' out a cone of lower \ion{H}{I} column density along which
Ly$\alpha$ photons escape. Studies of Ly$\alpha$ transfer through
simulated galaxies have also shown that the escape of Ly$\alpha$
photons from the dusty ISM of galaxies can proceed highly
anisotropically: the Ly$\alpha$ flux transmitted into different
directions may vary by an order of magnitude
\citep{Laursen09,Yajima12}. 

The {\it right panel} of Figure~\ref{fig:sb4} show the predicted azimuthally-averaged surface brightness \& polarization profiles
when the bipolar outflow is viewed from the top ({\it open circles}), and from the side ({\it open triangles}). The predicted polarization is lower (higher) when we view the outflow from the top (side), as photons typically scatter by $\mu < \cos 45^{\circ}$ ($\mu > \cos 45^{\circ}$). This panel also shows that the azimuthally averaged surface brightness profile depends only weakly on whether the outflow is bipolar or not, and on from which angle we view the outflow.
This last result is important as it can alleviate the problem that {\bf model IV} has, namely that $\sim 50\%$
of all Ly$\alpha$ photons do not scatter in the outflow, and should
thus be detectable as a Ly$\alpha$ source. Invoking this bipolarity
may help us avoid predicting a Ly$\alpha$ point source with Ly$\alpha$
halos, because when $\theta=45^{\circ}$, a fraction $\cos \theta \sim
0.71$ of all the sightlines would not lie in the cone, and we would
not see a point source along these sightlines. 

Two caveats to {\bf model V} are: ({\it i}) as mentioned and justified above, our model assumes that Ly$\alpha$ photons escape from the galaxy into the cones of outflowing gas. This `focussing' of Ly$\alpha$ photons into the biconical outflow requires additional \ion{H}{I} gas which is not present in our model. This gas would contribute to the measured EW at $b=0$, and would hence reduce the amount of \ion{H}{I} gas that we are allowed to assign to the clumps at small impact parameters. This reduced amount of cold gas would suppress the predicted amount of scattering -- and hence the predicted surface brightness -- at small impact parameters (which would agree better with the data); ({\it ii}) if the outflowing material were indeed all confided to two cones with opening angles $\theta=45^{\circ}$, then we fail to explain why blueshifted absorption lines of low ionization absorption lines are observed in practically all sightlines \citep[e.g.][]{Shapley03,Steidel10}. We intend to include the additional constraints provided by the `down--the--barrel spectra' in a more realistic model of the outflows, in which the clump distribution is not only a function of radius $r$, but also of azimuthal angle (see \S~\ref{sec:outlook}).

\section{Discussion}
\label{sec:discuss}

\subsection{Discussion of Model Uncertainties \& Caveats}
\label{sec:caveats}

Here, we discuss potential implications of our adopted assumptions and
simplifications.

\begin{itemize}

 \item We constrain the HI content of the cold clumps in the
   large-scale outflow with the mean observed absorption line strength
   (EW) as a function of impact parameter ($b$). These same clumps
   scatter Ly$\alpha$ photons emitted by the foreground galaxy into
   the line-of-sight, which could weaken the observed line strength
   (see Prochaska et al. 2011). For example, the observed surface
   brightness at $b=31$ kpc is $6 \times 10^{-19}$ erg s$^{-1}$
   cm$^{-2}$ arcsec$^{-2}$. The Ly$\alpha$ absorption strength has
   been measured over an area of $1.2 \times 1.35$ arcsec$^{2}$
   (C. Steidel, private communication), which implies that the
   observed absorption is accompanied by  $\sim 10^{-18}$ erg s$^{-1}$
   cm$^{-2}$ of scattered radiation. 
 
 We can compare this scattered flux to the flux that has been absorbed
 out of the spectrum of the background galaxy as follows: ({\it i})
 the observed flux density ($f_{\lambda}$ in erg s$^{-1}$ \AA$^{-1}$)
 at restframe $\lambda=1100$ \AA \hs is comparable to that at
 $\lambda=1500$ \AA\hs (see Fig~5 of Steidel et al. 2010, and using
 that $f_{\nu} \propto \lambda^2 f_{\lambda}$); ({\it ii}) we estimate
 the observed restframe flux density at $\lambda=1500$
 \AA\hs(restframe) from the average UV-derived star formation rate,
 uncorrected for dust, which was SFR$=8\hs M_{\odot}$ yr$^{-1}$
 assuming the Kennicutt (1998) star formation calibrator
 \citep{Erb}. For a galaxy at $z=2.3$ this star formation rate
 translates to an observed restframe $f_{\lambda}(\lambda=1500\hs{\rm
   \AA})\sim 2 \times 10^{-18}$ erg s$^{-1}$ cm$^{-2}$ \AA$^{-1}$.
 After combining ({\it i}) and ({\it ii}), we find that the scattered
 flux of $\sim 10^{-18}$ erg s$^{-1}$ cm$^{-2}$ corresponds to a rest
 frame EW$\sim 0.5$ \AA, which lies a factor of $4$ below the measured
 EW. This implies that scattered flux is sub-dominant to the absorbed
 flux. At larger impact parameters the scattered flux becomes
 exponentially fainter and even less important.

Obviously, our calculation is approximate and relies on an average
spectrum, an average star formation rate, and a median redshift. In
certain cases, we expect the scattered flux to be more important than
our previous estimate. However, in such cases the spatially extended
scattered flux is detectable in spectra of pixels adjacent to the
background galaxy, and can be corrected for. We therefore conclude
that the potential contribution of scattered radiation to the observed
EW does not affect our results.

\item When constraining the HI content of the clumps (using the
  observed EW as a function of $b$), we implicitly assume that only
  the cold clumps in the outflow contribute to the observed
  EW. However, \citet{VdV} have recently found that in their
  hydrodynamical simulations, the cold streams of gas that are feeding
  the central galaxy may produce more large column density absorbers
  ($N_{\rm HI} \gsim 10^{20}$ cm$^{-2}$) around galaxies than outflows
  (also see Fumugali et al. 2011, but note that modeling the
  outflowing component at large impact parameter is extremely
  uncertain, see \S~\ref{sec:clump2}). The probable contribution of
  cold streams to the observed EW at a given impact parameter reduces
  the amount of HI that we can assign to cold clumps in the
  outflow. With less HI in the outflowing clumps, we expect them to
  scatter fewer Ly$\alpha$ photons, and that consequently our surface
  brightness levels are reduced by an amount which depends on the the
  overall contribution of the outflowing cold clumps to the observed
  EW at a given $b$. 

\item The observed EW as a function of $b$ has been measured around
  galaxies with a mean redshift $\langle z \rangle =2.2$
  \citep{Steidel10}, while the mean redshift of galaxies that were
  used in the stack of narrowband images was $\langle z_{\rm halo}
  \rangle =2.65$. While galaxies in both samples were both selected in
  very similar ways, and therefore likely trace similar populations
  (e.g. both populations have virtually identical mean star formation
  rates, see \S~\ref{sec:cmod}), the observed Ly$\alpha$ absorption
  line strength does not really probe the same gas that is scattering
  the Ly$\alpha$ halos.  For this reason, we consider differences
  between the predicted and observed Ly$\alpha$ surface brightness
  profiles at the factor of 2-3 level (as was the case in {\bf model
    IV}) not a problem. On the other hand, it is unrealistic to
  attribute the order of magnitude differences in the predicted
  surface brightness profiles (which we found for {\bf model I}-{\bf
    model III}) to selection effects.

\item In our model in which the clumps decelerate after their initial
  acceleration (see \S~\ref{sec:vel}), the clumps reach a maximum
  velocity of $v_{\rm c, max}\sim 350$ km s$^{-1}$, which is well
  below the maximum velocity that was inferred by Steidel et
  al. (2010) of $v_{\rm max} \sim 800$ km s$^{-1}$. Fujita et
  al. (2009) found in their hydrodynamical simulations that a small
  fraction of the cold shell fragments could be accelerated to reach
  large outflow velocities ($\gsim 750$ km s$^{-1}$), while the bulk
  of the gas was traveling at lower velocities of $\sim 300$ km
  s$^{-1}$. This suggests that the maximum velocity that is inferred
  from the observations can be consistent with our model which only
  describes the kinematics of this bulk of the gas. Furthermore,
  because of the large HI column densities in the clumps, the photons
  can scatter in the wings of the Ly$\alpha$ line profile. As a
  result, Ly$\alpha$ absorption may trace a wider range of velocities
  than the range of actual outflow velocities. 

\item In our model there is a one--to--one mapping between radius and
  velocity. In reality, we expect outflows to have a range of
  velocities at a given radius: e.g. in the model of \citet{Martin05},
  the cloud acceleration increases with their HI column density. For a
  range of HI column densities we therefore expect the cold clouds to
  have a range of velocities at a given radius. This scatter may boost
  the predicted Ly$\alpha$ surface brightness profile,  in particular
  when this scatter gives rise to a population of clumps with a lower
  $v_{\rm c}(r)$.

\item Our model assumes that there is a unique clump mass. In reality,
  there is a distribution. This is very likely not an issue: in {\bf
    model I} the clump mass was $m_{\rm clump}\sim 10^4 M_{\odot}$,
  while in {\bf model II} the clump mass was $\sim 10$ times
  larger. As long as the distribution of clumps is constrained by the
  absorption line data, we predicted virtually identical surface
  brightness profiles. We therefore consider it unlikely that we
  obtain significantly different surface brightness profile, if we
  assume a finite range of clump masses.

\item Our model assumes a Gaussian emission profile, for which the
  width is set by the velocity dispersion of the gas. However, the
  outflow contains swept-up shells of cold neutral gas before it
  breaks out of the galaxy. It is therefore not unlikely that
  Ly$\alpha$ photons scatter off these dense cold shells, which would
  result in an overall redshift of the line, {\it before the
    Ly$\alpha$ photons escape from the galaxy into the large scale
    outflow}. Redshifting of the Ly$\alpha$ line would reduce the
  overall scattering probability in the outflow, and could reduce the
  predicted surface brightness profiles.

\item We assumed that the escape fraction\footnote{This escape
  fraction refers to the fraction of photons that escape from the ISM
  of the galaxy into the large-scale outflow. This escape fraction
  differs from the escape fraction that is used in recent
  observational papers \citep[e.g.][]{Hayes10,Hayes2,Blanc}, which
  represents the ratio of the observed to the intrinsic Ly$\alpha$
  flux, which can depend sensitively on the surface brightness
  threshold of the observations \citep[][also see Jeeson-Daniel et al. in prep]{ZZ10}.} of Ly$\alpha$ photons
  was 50-100\% (depending on the choice of IMF, and gas metallicity as
  these affect the {\it intrinsic} Ly$\alpha$ luminosity of the galaxy
  at a fixed SFR). This potentially high escape fraction was already
  noted by \citet{Steidel11}. In our models, it is required to be
  higher by a factor of $\sim 2$ because about half of the photons
  that escape from the galaxy into the large scale outflow were {\it
    not} scattered at all, and are effectively wasted. 

This radiation that is not scattered in the outflow must be a point
source of comparable ({\bf model IV}) or higher ({\bf model I }-{\bf
  model III}) luminosity than the halo itself, which is in conflict
with observations. This disagreement can be alleviated by invoking
that the outflow is bipolar (see \S~\ref{sec:bipolar}), and/or by a
population of low column density absorbers (as observed in
galaxy-quasar pair data by Rudie et al. 2012) that have a velocity
$v_{\rm c}$ that differs substantially from that in of the clumps in
{\bf model I-V} (see \S~\ref{sec:galqso}).

\end{itemize}

\subsection{Connection with Ly$\alpha$ Blobs}
\begin{figure}
\vbox{\centerline{\epsfig{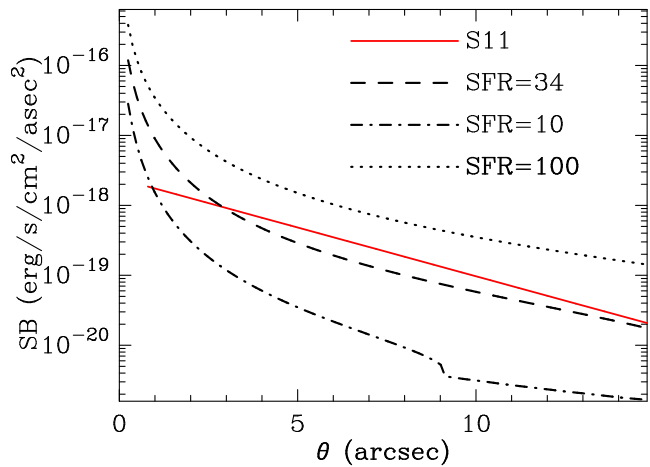}}}
\vspace{0mm}
\caption[]{This plot shows the dependence of our surface brightness
  profiles of the Ly$\alpha$ halos, as a function of the star
  formation rate (SFR, in $M_{\odot}$ yr$^{-1}$), assuming the same
  model parameters for the outflow (mass loading factor, number of
  clumps, etc.). The predicted surface brightness profile depends
  strongly on SFR because ({\it i}) the intrinsic Ly$\alpha$
  luminosity of the central source scales linearly with SFR, and ({\it
    ii}) the total amount of scattering material scales linearly with
  SFR. As a result, changing SFR by a factor of $\sim 3$  can change
  the surface brightness at a fixed $\theta$ by an order of magnitude} 
\label{fig:sfr}
\end{figure}

\citet{Steidel11} found that the Ly$\alpha$ `blobs' -- defined loosely
as having an area $\gsim 16$ arcsec$^{-2}$ for which SB$\gsim
10^{-18}$ erg s$^{-1}$ cm$^{-2}$ arscec$^{-2}$ (Matsuda et al. 2004,
also see Steidel et al. 2000, Saito et al. 2006) -- within their
narrowband survey volume to have surface brightness profiles almost
identical to that of the Ly$\alpha$ halos, apart from an overall
off-set in the surface brightness. Scattering of Ly$\alpha$ photons in
a large scale outflow can theoretically explain Ly$\alpha$ blobs, if
we increase the star formation rate of the central Ly$\alpha$
source. This is illustrated in Figure~\ref{fig:sfr}, which shows the
dependence of the predicted surface brightness profile on SFR for {\bf
  model IV} (which was also shown in Fig~\ref{fig:sb4}). In these
calculations we kept all other model parameters fixed.

The {\it dashed line} shows the original prediction for {\bf model IV}
which assumed SFR=34 $M_{\odot}$ yr$^{-1}$. The {\it dotted line}/{\it
  dot-dashed line} shows the predicted surface brightness profile for
SFR=100 $M_{\odot}$ yr$^{-1}$ / SFR=10 $M_{\odot}$ yr$^{-1}$. The
predicted surface brightness profile depends quite strongly on
SFR. This is because: ({\it i}) the intrinsic Ly$\alpha$ luminosity of
the central source scales linearly with SFR, and ({\it ii}) the amount
of cold gas that can scatter the Ly$\alpha$ photons also depends
linearly on SFR. These two effects combined suggest that the surface
brightness at a given impact parameter can depend on SFR as $\propto$
SFR$^{2}$, which can explain that the surface brightness at a given
impact parameter can vary by an order of magnitude as a result of a
factor of $\sim 3$ change on the SFR (which gives more weight to the
contribution of high-SFR galaxies to the stacked Ly$\alpha$
image). The feature in the SFR=10  $M_{\odot}$ yr$^{-1}$ curve at
$\theta\sim 9$ arcsec reflects that the clumps do not self-shield at
$r \gsim 76$ kpc in this model.

Figure~\ref{fig:sfr} shows that it is possible to have Ly$\alpha$
blobs around galaxies that are forming stars at a rate SFR$\gsim 100
M_{\odot}$ yr$^{-1}$, as the predicted surface brightness profile
drops below $\sim 10^{-18}$ erg s$^{-1}$ cm$^{-2}$ arcsec$^{-2}$ only
at $\theta \gsim 7-8$ arcsec. There is observational evidence that a
fraction of the Ly$\alpha$ blobs are associated with sub-mm galaxies,
which are believed to be forming stars at rates of SFR$\sim 10^3
M_{\odot}$ yr$^{-1}$ \citep{Chapman01,Geach05,Geach07,Matsuda07}.
\citet{Chapman05} have noted that a remarkably high fraction ($\sim
50\%$) of sub-mm galaxies show Ly$\alpha$ emission lines in their
spectra \citep[also see][for a detection of Ly$\alpha$ from a
  ULIRG]{NM09}. This may indeed suggest that enough Ly$\alpha$ photons
escape from the dusty interstellar media of sub-mm galaxies, and then
scatter in the large scale outflow to account for the blobs. Support
for the notion that Ly$\alpha$ blobs consist of scattered radiation is
provided by the detection of polarization by \citet{Hayes11}, albeit
at a level that is lower by a factor of $\sim 2$ than the values
predicted here. On the other hand, \citet{Prescott11} put an upper
limit on $P \lsim 10\%$ in LABd05 \citep{Dey05}, which clearly rules
out our models.

It appears increasingly plausible that there are distinct physical
mechanisms that power Ly$\alpha$ blobs. The polarization measurement
of \citet{Hayes11} clearly favors models that invoke scattering. On
the other hand, \citet{Prescott12} find that the UV continuum
(non-ionizing) associated with LABd05 is also spatially extended,
which favors having spatially extended emission of both UV and
Ly$\alpha$ photons. This observation could be consistent with {\it
  dust scattering}, which would explain why the polarization of
Ly$\alpha$ would not have been detected (see
\S~\ref{sec:lyamc}). However, given the large observed EW of the diffuse Ly$\alpha$ when measured relative to the diffuse UV-continuum \citep[REW$\gg200$ \AA,][]{Prescott12}, this is not very plausible. Alternative support for the
notion that some blobs are not a result of scattering is provided by
those blobs that do not have any clear galaxy counterparts
\citep[e.g.][]{Nilsson06,Smith07}.

\subsection{Additional Constrains from Galaxy-Quasar Pair Data}
\label{sec:galqso}

\citet{Rakic11} and \citet{Rudie12} use galaxy-quasar pairs to probe the CGM of the
foreground galaxies. \citet{Rudie12} present 10 pairs for which the background quasar
lies at $b < 100$ kpc. They find 6 absorbers with $\log N_{\rm
  HI}\gsim 17.0$, of which 2 absorbers have $\log N_{\rm HI}\gsim
18.0$, of which one absorber has $\log N_{\rm HI}\sim 20$ . We compare
this to {\bf model V} at $b=71$ kpc, which corresponds to the median
value for $b$ if the sightlines are distributed randomly within the
circle for radius $b=100$ kpc. We find that our model predicts $\sim
7$ absorbers with $\log N_{\rm HI} \gsim 17.0$, of which $\sim 7$ have
$\log N_{\rm HI} \gsim 18.0$, of which $\lsim 1$ absorber has $\log
N_{\rm HI} \gsim 19.3$. Given the simplified nature of our model, and
the relatively small number of observed sightlines, we consider these
numbers encouraging. For example, our predicted number of absorbers
with $\log N_{\rm HI} \gsim 18.0$ is reduced to $\sim 3$, and
therefore more consistent with observations, if we simply increase the
critical number density above which gas self-shields by a factor of
$2$ to $n_{\rm crit}=0.012$ cm$^{-3}$, which is still reasonable
(e.g. Faucher-Giguere et al. 2010 adopt $n_{\rm crit}=0.01$
cm$^{-3}$). Such a modification reduces the predicted EW as a function
of $b$, but only significantly at $b \gsim 70$ kpc. In this case the
observed EW at $b=100$ kpc could be accounted for by a large number of
lower column density absorbers, which have been observed (see below)
but which are not present in the model. 

Another difference worth emphasizing is that our model predicts that
ten sightlines with $b=71$ kpc should intersect a total of $\sim 13$
cold clumps. \citet{Rudie12} find significantly more low column
density, $\log N_{\rm HI}=14.5-17.0$, absorbers. This implies that our
clumpy outflow model does not account for all the observed absorbers,
and thus all potential `scatterers'. However, if we wish to use low
column density absorbers to scatter photons into Ly$\alpha$ halos,
then they must have lower outflow velocities (or they have to be
inflowing) than the clumps in our model, otherwise they are
transparent to the Ly$\alpha$ photons. The possible presence of these
low-column density that move at different velocities than the high
column density clumps in our models, may have the interesting benefit
that they reduce the fraction of photons that do not scatter in the
outflow at all. These clumps may thus reduce the luminosity of the
central Ly$\alpha$ sources that accompanies the Ly$\alpha$ halos in
{\bf model I-IV}, and in {\bf model V} when we view the outflow down
one of the cones.
 
\subsection{Outlook \& Potential Improvements}
\label{sec:outlook}

In \S~\ref{sec:caveats} we highlighted the simplifications of our
model, which underlined that many improvements are possible. We
discuss some examples of how we intend to improve upon our analysis
below.

 We have so far focused on using the observed Ly$\alpha$ absorption
 line strengths as well as the surface brightness profile of
 Ly$\alpha$ halos to constrain parameters of outflow models. However,
 there is information encoded in the observed spectral line shape of
 the Ly$\alpha$ emission line \citep[e.g.][and references therein]{Yamada12}, as well as whether it is redshifted or
 blueshifted relative to the galaxies systemic velocity. For example,
 Ly$\alpha$ lines that are blueshifted/redshifted with respect to the
 systemic velocity -- to first order -- are indicative of scattering
 through an opaque inflowing/outflowing medium
 \citep[e.g.][]{Zheng02,D06}. The first joint H$\alpha$ - Ly$\alpha$
 spectral line observations of spatially extended Ly$\alpha$ nebulae
 \citep{Yang11}, compact Ly$\alpha$ selected galaxies (Finkelstein et
 al. 2011, also see McLinden et a 2011 for joint
 Ly$\alpha$-[\ion{O}{III}] observations), and `double-peaked'
 Ly$\alpha$ emitting UV-selected galaxies \citep{Kulas12} have
 recently been reported. \citet{Kulas12} have already shown that such
 observations can rule out the `shell-models' for outflows. In the
 future we plan to explore what additional constraints we can place on
 outflow models on small scales (i.e. $r\lsim 10$ kpc) with
 observations of the Ly$\alpha$ spectral line shape (and
 shift). Finally, our previous discussion (\S~\ref{sec:galqso}) showed
 that observations of galaxy-quasar pairs \citep{Rakic11,Rudie12} already
 provide useful additional constraints on our models.

We also plan to extend our study to include other lines. For example,
\citet{Bordoloi11} have presented the radially (and azimuthally)
dependent absorption line strength of \ion{Mg}{II} around bright
flux-selected galaxies at $0.5 < z< 0.9$ from the z-Cosmos redshift
survey. We can constrain the \ion{Mg}{II} content of the clumps in our
model by matching this data. We can then make predictions for surface
brightness profiles of scattered \ion{Mg}{II} emission, and compare to
observations of spatially extended \ion{Mg}{II} emission around a
$z=0.69$ starburst galaxy \citep{Rubin11}.

\section{Conclusions}
\label{sec:conc}
We have presented `constrained' radiative transfer calculations of
Ly$\alpha$ photons propagating through clumpy, dusty, large scale
outflows, and explore whether scattering through such an outflow can
quantitatively explain the Ly$\alpha$ halos that have been observed
around Lyman Break Galaxies (LBGs, see Steidel et al 2011). As part of
our analysis we have modified a Ly$\alpha$ Monte-Carlo radiative
transfer code to allow us to follow the propagation of Ly$\alpha$
photons through a multiphase, dusty medium for arbitrary distributions
of clumps. This code also computes the polarization of the scattered
Ly$\alpha$ radiation. Previous calculations of the polarization of
scattered Ly$\alpha$ radiation have focused only on homogeneous
spherically symmetric gas clouds or shells. We have successfully
tested our code against several analytic solutions, some of which --
in particular the directional dependent frequency redistribution
function -- are new.

 Modeling the distribution and kinematics of cold gas in outflows from
 first principles is an extremely complex task, which likely requires
 magneto-hydrodynamical simulations that have sub-pc resolution
 (\S~\ref{sec:clump2}). We have taken an different approach, and
 constructed phenomenological models for the large-scale outflows in
 which cold ($\log T_{\rm c} \sim 3-4$) clumps are in pressure
 equilibrium with a hot ($\log T_{\rm h} \sim 7$) wind. We first
 considered models in which the cold clumps are distributed
 symmetrically around the source, and which accelerate continuously as
 they break out of the interstellar medium of the
 galaxy. \citet{Steidel10} showed that this type of model may
 qualitatively simultaneously explain the observed Ly$\alpha$
 absorption line strength in the CGM, as well as the observed surface
 brightness profiles of Ly$\alpha$ emission line halos.
  
 Our more detailed analysis shows that such models-- which contain
 $10^{5-6}$ discrete clumps--can reproduce the observed Ly$\alpha$
 absorption strength of the circumgalactic medium measured in the
 spectra of background galaxies very well, and for model parameters
 that are physically plausible (see Fig~\ref{fig:ewb}).  However, when
 we insert a Ly$\alpha$ source in the center of these clumpy outflow
 models,  and compute the observable properties of the scattered
 Ly$\alpha$ radiation, we typically find that the predicted
 Ly$\alpha$ halos are significantly fainter and more concentrated than
 what is observed (Fig~\ref{fig:sb1}). The reason for this discrepancy
 is easy to understand: outflowing cold clumps that scatter photons at
 large ($b \gsim 30$ kpc) impact parameters are propagating away from
 the Ly$\alpha$ source at $v \gsim 600$ km s$^{-1}$. In order for the
 clumps to scatter Ly$\alpha$ photons they must be opaque to these
 photons, which requires HI column densities in excess of $N_{\rm HI}
 \gsim 10^{19}$ cm$^{-2}$. However, the absorption line data requires
 that the number of such clumps at large impact parameters is so small
 that a significant fraction of the photons never encounter them. Our
 conclusion that we cannot simultaneously fit the absorption line and
 Ly$\alpha$ halo data --with a clumpy outflow in which the clumps are
 distributed spherically around the galaxy and which accelerate with
 radius -- is therefore robust.
 
 We also found that the vast majority of photons scatter in zero or
 one clumps, and that it is possible to analytically compute the
 Ly$\alpha$ surface brightness and polarization profiles (see
 Fig~\ref{fig:sb1}).  The fact that a significant fraction of the
 photons do not scatter in the outflow is problematic. The photons
 that do not scatter must be detectable as a point source, and it's
 predicted luminosity equals or exceeds the total luminosity of the
 Ly$\alpha$ halo, which is in further disagreement with the
 observations. 
   
 We can much better simultaneously reproduce the observed Ly$\alpha$
 absorption line strengths and the Ly$\alpha$ halos with models in
 which the cold outflowing clumps decelerate (see
 Fig~\ref{fig:sb4}). This deceleration occurs naturally in models of
 momentum-driven winds (\S~\ref{sec:vel}). We can alleviate the
 problem of predicting a bright Ly$\alpha$ point source to accompany
 the halo, if the outflow is bipolar with an (half) opening angle
 $\theta \lsim 45^{\circ}$ (\S~\ref{sec:bipolar}). This problem may be
 further reduced if the observed additional low-column density
 absorbers in galaxy-quasar pairs move at velocities that allow them
 to resonantly scatter an additional fraction of the Ly$\alpha$
 photons (see \S~\ref{sec:galqso}).
 
We found that models which do fit both the absorption line strength
and Ly$\alpha$ halo data give rise to levels of linear polarization
that reach $P\sim 40\%$ at a surface brightness level of SB$\sim
10^{-18}$ erg s$^{-1}$ cm$^{-2}$ arscec$^{-2}$
(e.g. Fig~\ref{fig:sb4}). This polarization signature is likely unique
to the scattering models and likely distinguishes them from models in
which the Ly$\alpha$ photons were emitted over a spatially extended
region (see \S~\ref{sec:intro}). Furthermore, because the large
polarization signature is a result of  non-resonant scattering, the
polarization also distinguishes our models from those of \citet{ZZex},
in which the halos were a result of resonant scattering
(\S~\ref{sec:intro}). It should be noted that it remains to be shown
that predictions of these other models are in quantitative agreement
with the observed Ly$\alpha$ absorption line data, as well as the
Ly$\alpha$ halos.

This paper illustrates clearly that Ly$\alpha$ emission line halos
around star forming galaxies provide valuable constraints on the cold
gas distribution \& kinematics in their circumgalactic medium, and
that these constraints nicely complement those obtained from
absorption line studies alone.

{\bf Acknowledgments} We thank Alice Shapley, Andrea Ferrara, Akila
Jeeson-Daniel, Benedetta Ciardi, Evan Scannapieco, Rongmon Bordoloi,
for helpful discussion. We thank Charles Steidel for helpful
correspondence, and Matthew Hayes \& Moire Prescott for helpful comments on an earlier
version of this paper. We thank an anonymous referee for a prompt report, and for carefully reading our paper.

\label{lastpage}

\appendix

\section{Code Description \& Testing}
\begin{figure}
\vbox{\centerline{\epsfig{file=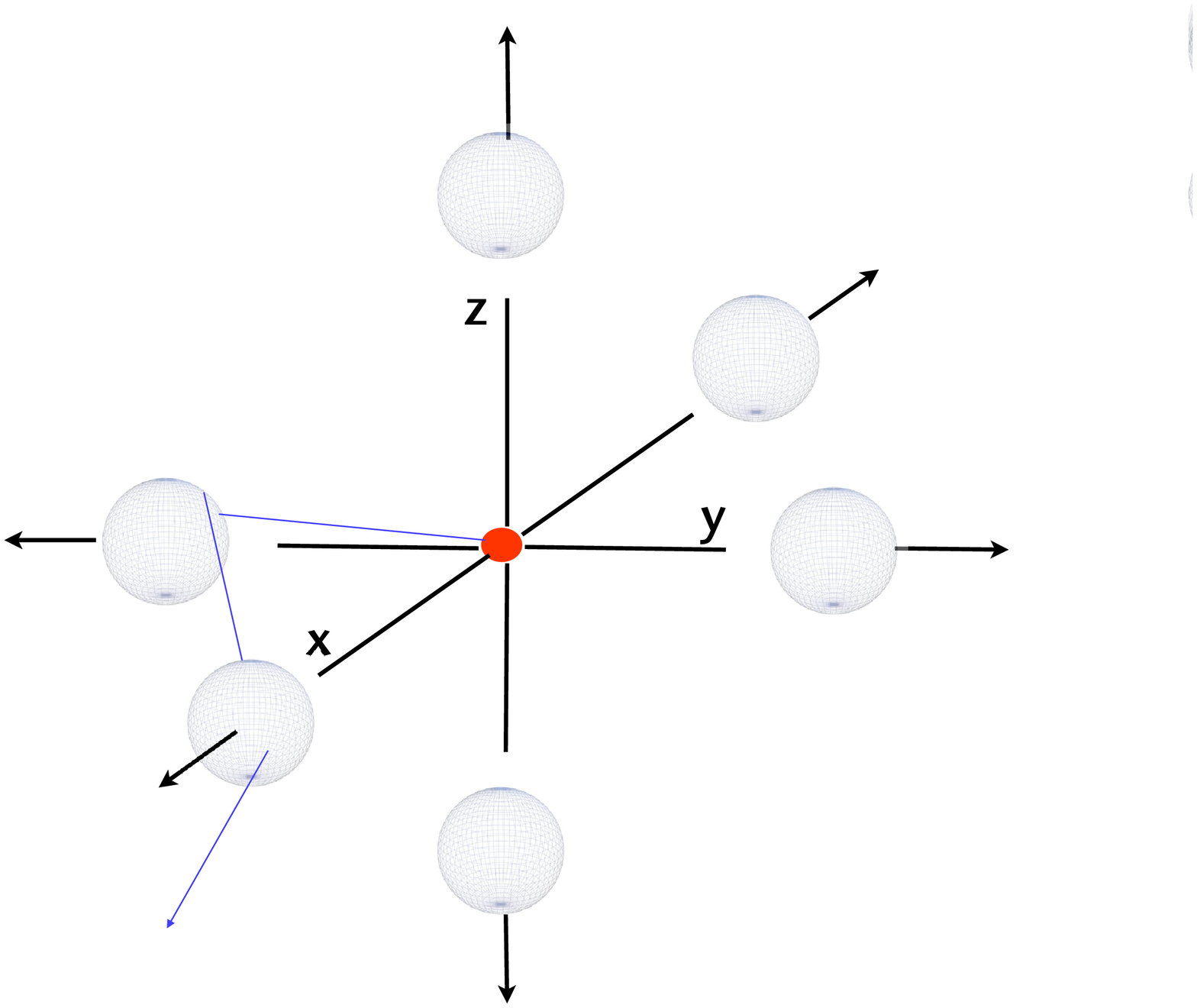,angle=270,width=9.5truecm}}}
\vspace{-2mm}
\caption[]{Clump geometry that we assumed in our test
  calculations. Six clumps of radius $R_c$ lie on the coordinate axes
  a distance $d=50$ kpc from the origin, which contains the source of
  Ly$\alpha$ photons. When the clumps are optically thin to Ly$\alpha$
  photons, we can analytically compute the fraction of photons that
  scatter in $N$ clumps as a function of $N$ (see
  \S~\ref{app:thin}). When the clumps are opaque to Ly$\alpha$ photons
  (see \S~\ref{app:thick}), we can analytically compute the emerging
  spectrum.} 
\label{fig:app2}
\end{figure} 
\begin{figure*}
\vbox{\centerline{\epsfig{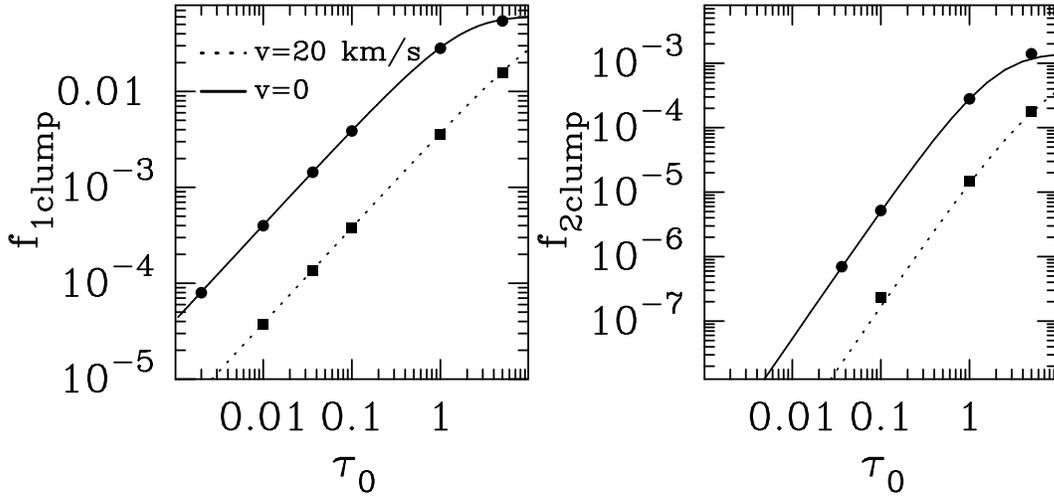}}}
\vspace{-2mm}
\caption[]{Comparison between analytic and Monte-Carlo calculations of
  the fraction of photons that scatter in one clump ({\it left panel})
  and in two {\it distinct} clumps ({\it right panel}) as a function
  of the line center optical depth of an individual clump, $\tau_0$,
  for $v_{\rm c}=0$ ({\it solid lines}), and $v_{\rm c}=20$ km
  s$^{-1}$ ({\it dotted lines}). The agreement is excellent, and shows
  that inter and intra-clump propagation--as well as frequency
  redistribution by moving clumps--are captured well by the
  Monte-Carlo code.} 
\label{fig:app3}
\end{figure*}

We describe modifications to the code of \citet{D06} in detail. We
test various subroutines of the new code in a simple geometry in which
six clumps of radius $R_c=10$ kpc lie on the coordinate axes a
distance $d=50$ kpc from the origin. That is, the clumps lie at  $(x
=\pm d,0,0)$, $(0,y =\pm d,0)$, and $(0,0,z=\pm d)$, where $d=50$ kpc.
We assign an outflow velocity $v_{\rm c}$ to the clumps

In section \S~\ref{app:thin} we consider cases in which the clumps are
transparent (i.e. $\tau \ll 1$) to Ly$\alpha$, and in
\S~\ref{app:thick} we consider cases in which the clumps are extremely
opaque to Ly$\alpha$ photons. For all these tests we assume that the
gas temperature of the gas in the clumps is $T_{\rm c}=10^4$ K.

\subsection{Central Source \& Transparent Clumps}
\label{app:thin}

For these tests, we insert all Ly$\alpha$ photons at the origin, and
at the line resonance, i.e. $x_{\rm in}=0.0$.

\subsubsection{Testing the Interclump Propagation Scheme}
The sky covering factor of a single clump for a central source is
$f_{\rm c}=\frac{\Omega_{\rm c}}{4\pi}$, where $\Omega_{\rm c}= \pi
\theta_{\rm c}^2$ in which $\theta_{\rm c} =$ arcsin$(R_c/d)$. The
fraction of photons that scatter in only one clump\footnote{Formally,
  we have to multiply this probability $f_{\rm 1scat}$ by the
  probability that photons do {\it not} subsequently scatter in other
  clumps. As the clump sky covering factor is only $\sim 6\%$, this
  introduces a correction of at most a factor of $\sim
  (1.0-0.06)=0.94$.} is $f_{\rm 1clump}=6 \times f_{\rm c} \times
\langle1-  \exp[ -\tau(x_{\rm in})]\rangle$. Here $\langle 1- \exp
[-\tau(x_{\rm in})]\rangle$ is given by
\begin{eqnarray}
\langle 1- \exp [-\tau(x_{\rm in})]\rangle = \\ \nonumber \frac{2
  \pi}{\pi R^2} \int_0^R ydy\hs \Big{(}1-\exp\Big{[} -\tau_0 \times
  \sqrt{\frac{R^2-y^2}{R^2}} \phi(x'[y])\Big{]}\Big{)}.
\label{eq:trans1}
\end{eqnarray} Here $y$ denotes the impact parameter from the center of the clump to where the photon strikes, $\tau_0$ denotes the line center optical depth through the center of the clump, and $\phi(x'[y])$ denotes the Voigt profile evaluated at $x'=x_{\rm in}-\frac{v_{\rm c}}{v_{\rm th}}\cos \theta$, where $\theta$ denotes the angle between the photon's wavevector ${\bf k}$ and the outflow velocity vector ${\bf v}$, and we have $\cos \theta =\sqrt{1 -\sin^2 \theta}=\sqrt{1-(y/d)^2}$.

For example, for stationary clumps, the $y-$dependence of the term
$\phi(x'[y])$ vanishes and the integral can be evaluated analytically
when $\tau_0 \ll 1$, resulting in $f_{\rm 1clump}\approx 0.0405 \tau_0
\phi(x)$. The {\it left panel} of Figure~\ref{fig:app3} compares
analytically computed values of $f_{\rm 1clump}$ for $v_{\rm c}=0$
({\it solid line}), and $v_{\rm c}=20$ km s$^{-1}$ ({\it dotted
  lines}) as a function of $\tau_0$, with those obtained from the
Monte-Carlo code. The agreement is excellent, and shows that inter and
intra-clump propagation are captured well by the Monte-Carlo code.

In the optically thin limit, we can also estimate the fraction of
photons that scatter in two different clumps. An accurate estimate for
this probability is given by 
\begin{eqnarray}
f_{\rm 2clump}=f_{\rm 1clump}\times \sum_{i \neq j}^{N_{\rm
    clump}}\frac{\Omega_{\rm c,i}}{4 \pi} P(\mu_i)\times \nonumber
\\ \int_{-\infty}^{\infty}dx' \hs \Big{[} \langle 1- \exp -\tau_{\rm
    i}(x')\rangle \Big{]}R(x'|x [x_{\rm in}],\mu_i).
\end{eqnarray} This equation gives the probability that a photon scatters in a second clump, denoted with number `i', {\it after having scattered in the first clump}, denoted with `j'. The probability that the photon scatters in a second clump is a product of the probabilities that the photon scatters into a sightline that intersects clump `i', and that it then scatters in that clump. This latter probability depends on the frequency of the photon after the first scattering event, and we integrate over all possible photon frequencies weighted by PDF of this frequency. A more detailed quantitative explanation follows  below.

Firstly,  the probability that the photon scatters into a sightline
that intersects clump `i' is given by $\approx \frac{\Omega_{\rm
    c,i}}{4 \pi} P(\mu_i)$. In our testcase,  a photon has to scatter
either by $\mu\approx -1$ for scattering in the clump on the same
coordinate axis ($i=5$), or $\mu \approx -1/\sqrt{2}$ for scattering
in clumps on one of the other coordinate axes ($i=1-4$). These
probabilities are approximations -but accurate ones- because in
reality the photons scatter into a (narrow) range of $\mu$, which in
detail depends on where exactly the first scattering event
occurred. To capture this effect properly, we would have to average
over $\mu$ weighted by the proper PDF for $\mu$. However, since this
range of $\mu$ only extends over $\Delta \mu \approx 0.2$, and because
both $P(\mu)$ and $R(x'|x[x_{\rm in}],\mu_i)$ change very little over
this range, this more detailed and tedious procedure barely changes
our final results.

Secondly, the expression for the probability that a photon scatters in
the second clump is given by  $\langle 1- \exp -\tau_{\rm
  i}(x')\rangle$, which is given by Eq~\ref{eq:trans1} for $i=5$ (with
$d=100$ kpc), but for $i=1-4$ we omit the $y-$dependence of $x'$. The
geometry for scattering in clumps $i=1-4$ does not allow for a simple
mapping between impact parameter $y$ and Doppler boost, and this last
modification represents a reasonable approximation.

Finally, the PDF for the outgoing photon frequency $x'$ depends on
both the scattering direction {\it and} incoming photon frequency
$x_{\rm in}$. We derive an analytic solution for the frequency
redistribution function, $R(x''|x,\mu)$, which denote the PDF for
$x''$ given $x$. These frequencies are measured in the frame of the
gas. In our test case, photons appear at frequency $x=x_{\rm
  in}-v_{\rm c}/v_{\rm th}$ in the frame of the first clump. The
outgoing photon frequency $x'$ (measured in the lab frame) relates to
$x''$ through a Lorentz transformation: $x'=x''+\mu v_{\rm c}/v_{\rm
  th}$. The expression for $R(x''|x,\mu)$ is derived in
Appendix~\ref{app:redist}.

The {\it right panel} of Figure~\ref{fig:app3} compares the
analytically computed values of $f_{\rm 2clump}$ as a function of
$\tau_0$ for $v_{\rm c}=0$ ({\it solid line}), and $v_{\rm c}=20$ km
s$^{-1}$ ({\it dotted lines}) with those obtained from our Monte-Carlo
code. The agreement is again excellent. This further illustrates that
inter and intra-clump propagation of photons is described
accurately. Furthermore, the (directional dependent) frequency
redistribution is also captured accurately.

\begin{figure*}
\vbox{\centerline{\epsfig{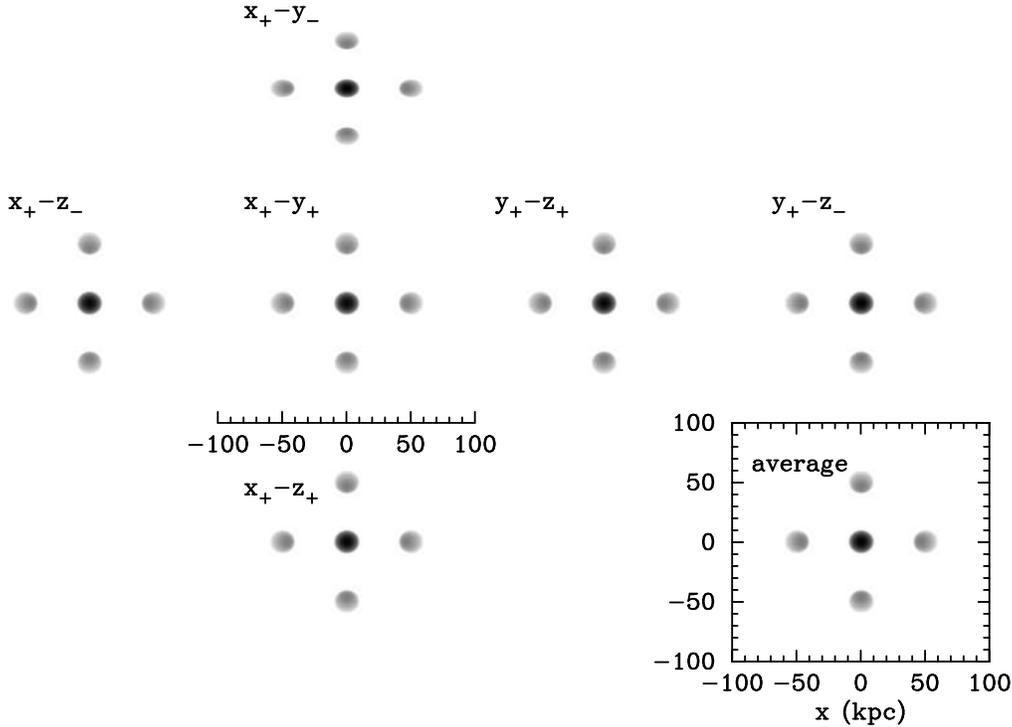}}}
\vspace{-2mm}
\caption[]{The surface brightness profiles for the scattered radiation
  are shown for all 6 viewing directions for a case in which
  $\tau_0=0.1$. The {\it lower right panel} shows the average surface
  brightness profile. The central clump is approximately twice as
  bright as this contains emission that scattered from two
  clumps. This plot illustrates nicely that all scattered radiation is
  confined to regions associated with clumps. A more detailed view of
  a the surface brightness profile of a single clump in the averaged
  image is shown in Figure~\ref{fig:cont2} } 
\label{fig:cont1}
\end{figure*}
\subsubsection{Testing the Surface Brightness \& Polarization Subroutines}
\label{app:nee}

\citet{D06} focused on spherically symmetric gas distributions, which
makes the calculation of the predicted surface brightness distribution
straightforward. In three-dimensional geometries ideally one has to
compute the fraction of photons that escapes from the medium, exactly
in the direction of the telescope. Since the telescope is a
cosmological distance removed from the location of last scattering,
and hence practically subtends an infinitesimally small fraction of
the sky, this procedure is not possible in practice.

We follow a standard approach for computing surface brightness
profiles, and compute the differential probability that a photon is
scattered exactly towards the telescope, for each scattering
event. This probability can be computed as follows. A photon scatters
at some location ${\bf x}_{\rm scat} \equiv (x_{\rm s},y_{\rm
  s},z_{\rm s})$, by an atom whose velocity components are given by
${\bf v}\equiv (v_{\rm x},v_{\rm y},v_{\rm z})$, or by a dust grain
whose thermal motion we neglect. We denote the photon's propagation,
frequency, and polarization before scattering with ${\bf k_{\rm in}}$,
$x_{\rm in}$, and ${\bf e_{\rm in}}$. Now let us consider the
$x^+-y^+$ image. Photons that make up this image would have to be
scattered into direction ${\bf k}_{\rm out}$=$(k_{\rm x},k_{\rm
  y},k_{\rm z})=(0,0,1)$. 

The probability per sterradian that a photon escapes into direction
${\bf k}_{\rm out}$ is

\begin{eqnarray}
P=\exp[-\tau(x_{\rm out},{\bf k}_{\rm out},{\bf x}_{\rm out})]\times
\\ \nonumber \times P({\bf k}_{\rm in},{\bf k}_{\rm out}, {\bf e}_{\rm
  in} | {\rm wing/res/dust}),
\end{eqnarray} where $x_{\rm out}$ is determined fully by $k_{\rm out}$, $x_{\rm in}$ and the velocity vector of the scattering atom, $v$ (see Eq~\ref{eq:xoxi1}), and where $P({\bf k}_{\rm in},{\bf k}_{\rm out}, {\bf e}_{\rm in} | {\rm wing/res/dust})$ denotes the phase function, which depends on whether the photon scatters in the line resonance, in the wings of the line, or off a dust grain. For wing scattering the phase function depends on ${\bf e}_{\rm in}$ as $P({\bf k}_{\rm out},{\bf e}_{\rm in}|{\rm wing})= \frac{3}{2}[1-({\bf k}_{\rm out}\cdot {\bf e}_{\rm in})^2]$ \citep{RL99,DL08}.

The total flux that the photon then contributes to the relevant pixel
on the image - in this case at $(x_{\rm s},y_{\rm s})$, is
given\footnote{The standard factor of $4 \pi$ is sometimes missing
  from the denominator in the literature
  \citep[e.g. in][]{Ta06,Laursen09}, because of the normalization of
  the phase-functions. Our phase functions are normalized as $\int
  d\Omega P({\bf k}_{\rm in},{\bf k}_{\rm out}, {\bf e}_{\rm in})
  \equiv 4 \pi$.} by $S=\frac{\mathcal{L}}{4\pi d^2_{\rm L}(z)}\times
P$, where $\mathcal{L}=L_{\rm tot}/N_{\gamma}$. Here, $L_{\rm tot}$
denotes the total Ly$\alpha$ luminosity of the source, and $N
_{\gamma}$ denotes the total number of photons used in the Monte-Carlo
run \citep[also see][]{Ta06,Laursen07}. 

Following \citet{RL99} the linear polarization $\mathcal{P}$ at a
given location is determined by the {\it polarized} fluxes $S_{\rm l}$
and $S_{\rm R}$ as
\begin{equation}
\mathcal{P} \equiv \frac{S_{\rm r}-S_{\rm l}}{S_{\rm r}+S_{\rm l}}.
\label{eq:pol1}
\end{equation} The total contribution of the photon to these polarized fluxes is given by 

\begin{eqnarray}
S_{\rm r}=&S \times \big{[}g(\mu)(1-\cos^2
  \chi)+\frac{1}{2}(1-g(\mu))\big{]}, \\ \nonumber S_{\rm l}=&S \times
\big{[} g(\mu) \cos^2\chi +\frac{1}{2}(1-g(\mu))\big{]},
\label{eq:appol}
\end{eqnarray} when a Ly$\alpha$ photon is scattered by a hydrogen atom. In this expression, $g(\mu)=1$ for wing scattering, and $g(\mu)=\frac{1+\mu^2}{11/3+\mu^2}$ for resonant scattering \citep{RL99,DL08}. Furthermore, $\chi$ denotes the angle between the photonÕs polarization vector {\it after scattering}, denoted with ${\bf e}_{\rm out}$, and the the vector ${\bf x}_{\rm scat}$ projected onto the $x-y$ plane \citep[as in][]{RL99,DL08}. In the case of wing scattering, ${\bf e}_{\rm out}$ is obtained by finding the normalized projection of the old polarization vector ${\bf e}_{\rm in}$ onto the plane normal to ${\bf k}_{\rm out}$ \citep{RL99}. In the case of resonant scattering, we generate a random unit vector perpendicular to ${\bf k}_{\rm out}$.  Eq~\ref{eq:pol1} shows for example that scattering by 90$^{\circ}$ (i.e. $\mu=0$) results in $\mathcal{P}=1.0$ for wing scattering, and  $\mathcal{P}=3/11$ for resonant scattering. When a Ly$\alpha$ photon scatters off a dust grain, we simply set $S_{\rm l}=S_{\rm r}=S/2$. 

\begin{figure}
\vbox{\centerline{\epsfig{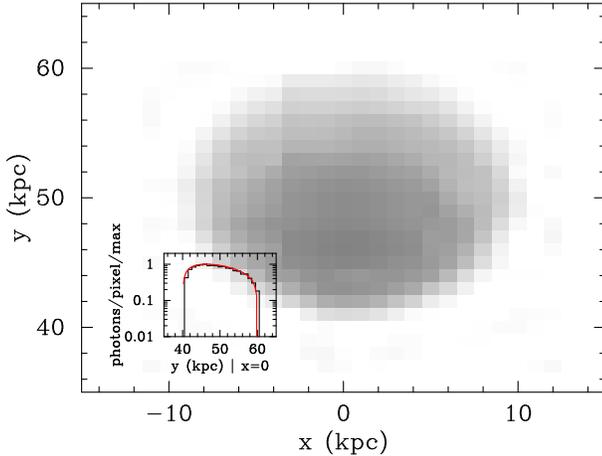}}}
\vspace{-2mm}
\caption[]{A more detailed view of the surface brightness distribution
  in the uppermost clump in the {\it averaged} image (shown previously
  in Fig~\ref{fig:cont1}) is shown. The inset shows a slice through
  this distribution along the axis $x=0$. The {\it black histogram}
  shows the normalized (to the maximum) surface brightness as a
  function of position $y$, which matches our analytic solution ({\it
    red solid line}) very well. This demonstrates that our surface
  brightness algorithm works well.} 
\label{fig:cont2}
\end{figure}

Figure~\ref{fig:cont1} shows the six images that we created from six
directions (along the $\pm \hat{x}$, $\hat{y}$, and
$\hat{z}$--directions), for a test problem in which $\tau_0=0.1$, and
$v_{\rm c}=0$ (see above). Because of our adopted geometry, the six
images look identical within the noise as a result of the finite
number of photons in the Monte-carlo run. The {\it lower right panel}
shows the image after taking the average of all
six. Figure~\ref{fig:cont2} shows a close-up of the upper most
clump. This figure demonstrates that our images contain no flux where
there is not supposed to be any. Furthermore, the surface brightness
profile of individual clumps is in good agreement with analytic
estimates: the {\it inset} shows a slice through the surface
brightness map at $x=0$ and plots the surface brightness --normalized
to the maximum surface brightness in the clump-- as a function of
$y$. The {\it histogram} shows the result obtained from our
Monte-Carlo code, while the {\it red solid line} shows our analytic
estimate, which we compute as follows:

The total flux that we expect to detect is $S(y) \propto \int
_{-l(y)}^{l(y)} ds\hs \tau(y,s)f(s,y)$. Here, $f(y,s)\propto
(y^2+s^2)^{-1}$ denotes the incoming flux at position $(y,s)$, where
$s$ denotes the position along the line of sight. This flux intersects
the line of sight at an angle that increases with $s$, and the
probability that the photon is scattered scales as $\tau \propto
\sqrt{y^2+s^2}/y$. We can therefore write $S(y) \propto
\frac{1}{y}\int _{-l(y)}^{l(y)} ds\hs (y^2+s^2)^{-1/2}$. At a given
$y$, we know we will exit from the clump at $\pm
l(y)=(R^2-(d-y)^2)^{1/2}$. The resulting $S(y)$ -normalized to its
maximum- is overplotted as the {\it red solid line}. The agreement
between our analytic and Monte-carlo calculations is excellent which
further confirms that or surface brightness algorithm is working well.

We also found that the linear polarization lies in the range
$\mathcal{P}=23-27\%$ when measured across the 4 clumps that are not
at $(x,y)=(0,0)$ in the averaged image. This is very close to the
maximum linear polarization, $\mathcal{P}_{\rm max,res}=\frac{3}{11}$,
that is expected from resonantly scattered Ly$\alpha$ radiation. The
polarization of radiation coming from the central clumps is consistent
with 0 in the center and rises to $\mathcal{P}\sim 1\%$ on the edges,
which is again consistent with analytical expectations, which yield
$\mathcal{P}=\frac{1-\mu^2}{11/3 +\mu^2}$\citep{DL08}. For scattering
in the central clumps we have $\sin \theta \approx \tan \theta \approx
y/d$, and the maximum polarization for the central clump is $\approx
\frac{1}{25}\times \frac{3}{14} \approx 0.9\%$, which is in close
agreement with our Monte-Carlo calculations.

\subsection{Extremely Opaque Clumps with Embedded Ly$\alpha$ Sources}
\label{app:thick}
\begin{figure}
\vbox{\centerline{\epsfig{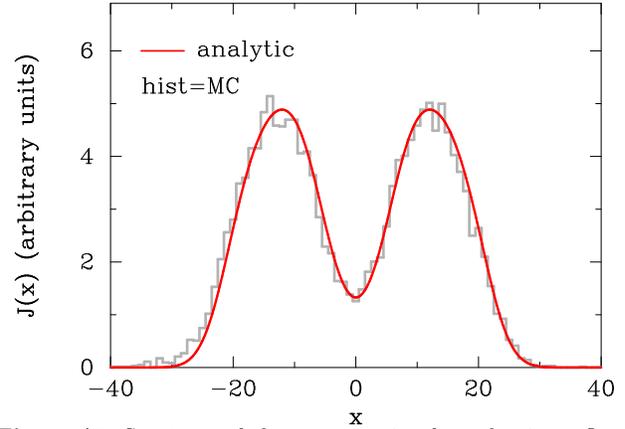}}}
\vspace{-2mm}
\caption[]{Spectrum of photons emerging from the six outflowing,
  optically thick clumps. Photons were emitted in the centers of all
  clumps. The {\it red solid line} shows the analytic solution
  (Eq~\ref{eq:thick2}), while the {\it histogram} shows the solution
  that we obtain from our Monte-Carlo simulation. This plot
  illustrates that our code treats scattering in the wing
  accurately. } 
\label{fig:thick}
\end{figure}
The previous sections showed that our code works well in the optically
thin regime. These tests were important as they demonstrates the
accuracy of our code with respect to resonant scattering, inter and
intraclump propagation, the surface brightness and polarization
algorithms. In this section we briefly describe one test of scattering
in an extremely opaque medium. This tests scattering in the wings of
the line in greater detail. 

We insert photons at line center in the centers of all clumps. The
line center optical depth through the clumps is enhanced to
$\tau_0=10^7$. The spectrum of photons emerging from individual clumps
is known analytically \citep{Harrington73,Neufeld90,D06} and is given
by
\begin{equation}
J_{\rm sph}(x)=\frac{\pi}{a_v\tau_0\sqrt{24}}\frac{x^2}{1+{\rm
    cosh}\hs{\Big{[}}\sqrt{\frac{2\pi^4}{27}}\frac{|x|^3}{a_v\tau_0}\Big{]}}.
\label{eq:dij}
\end{equation} This frequency $x$ is measured in the frame of the clump. When a clump is outflowing, then the proper Doppler boost should be applied. Under the -reasonable- assumption that a negligible fraction of the photons scatters in a second clump, then the total spectrum of photons emerging from the six outflowing clumps

\begin{equation}
J(x)=\frac{1}{2\Delta x}\int_{x-\Delta x}^{x +\Delta x} dx' \hs J_{\rm
  sph}(x')
\label{eq:thick2}
\end{equation}, where $\Delta x = v_{\rm c}/v_{\rm th}$. The {\it red solid line} shows Eq~\ref{eq:thick2} for $\tau_0=10^7$, and $v_{\rm c}=100$ km s$^{-1}$, while the {\it histogram} shows the spectrum of photons that escapes from out Monte-carlo simulation. This plot illustrates that our code treats scattering in the wing accurately. Further tests of our code regarding scattering in extremely opaque media were presented in \citet{D06} and \citet{DL08}.

\subsection{Dust Scattering \& Absorption}

\begin{figure}
\vbox{\centerline{\epsfig{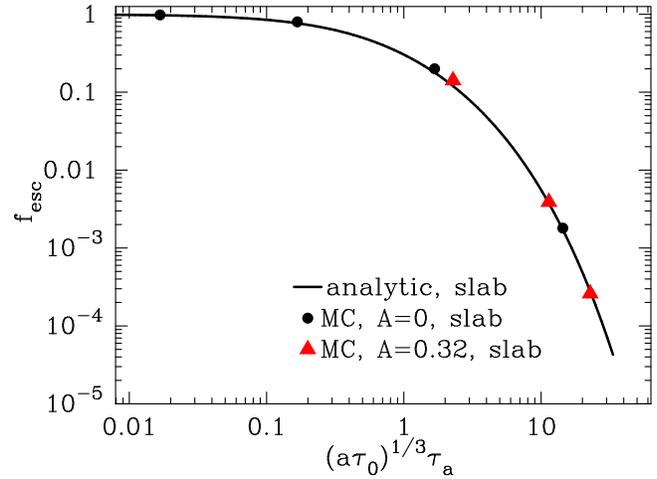}}}
\vspace{0mm}
\caption[]{The escape fraction, $f_{\rm esc}$, of Ly$\alpha$ photons
  from a uniform slab as a function of dust {\it absorption} opacity,
  $\tau_{\rm a} \equiv (1-A)\tau_{\rm d}$,  from the midplane to the
  edge of the slab.  Here, $\tau_{\rm d}$ denotes the {\it total} dust
  opacity, and $A$ denotes the assumed albedo of the dust grains. The
  {\it solid line} shows the analytic solution of
  \citet{Neufeld90}. The {\it black filled circles} ({\it red filled
    triangles}) denote the escape fraction obtained from our
  Monte-Carlo code assuming $A=0$ ($A=0.32$). Destruction of
  Ly$\alpha$ photons is captured accurately in both cases of zero and
  non-zero Albedo.}  
\label{fig:esc}
\end{figure}

Analytic expressions for the fraction of photons that escape from
uniform slabs (infinite plane parallel media) have been derived by
\citet{Harrington73} and \citet{Neufeld90}. \citet{Neufeld90} has
provided an approximate expression for the escape fraction of
Ly$\alpha$ photons for the case in which photons are emitted at line
center in the midplane of an `extremely opaque' slab, where `extremely
opaque' quantitatively means $a_v \tau_0 \geq 10^3$. This
expression\footnote{The numerical factor given in \citet{Neufeld90}
  also applies to our calculations, despite the different definition
  of $\tau_0$. This is because the original approximate form derived
  by \citet{Neufeld90} is given by$f_{\rm esc}=1/{\rm
    cosh}(Y_0^{1/2})$, where $Y_0=[3 \beta \phi(x_{\rm
      s})]^{1/2}\tau_0$. Here, $\beta \equiv (1-A)\tau_{\rm
    d}/\tau_0$, and $x_{\rm s}\equiv 0.525(a\tau_0)^{1/3}$. If we
  properly rescale $\tau_{0,{\rm neufeld}}\rightarrow
  \sqrt{\pi}\tau_{0,{\rm us}}$, and $\phi(x)_{{\rm
      neufeld}}\rightarrow \phi(x)_{{\rm us}}/\sqrt{\pi}$, then we get
  back the original equation.} is

\begin{equation}
f_{\rm esc}= \Big{[}{\rm cosh}
  \big{(}3.46(a_v\tau_0)^{1/3}(1-A)\tau_{\rm d}\big{)}^{1/2}
  \Big{]}^{-1}
\end{equation}  where $\tau_d$ is the {\it total} (absorption + scattering) optical depth in dust from the midplane to the edge of the slab, and where $A$ denotes the albedo. Figure~\ref{fig:esc} shows that the escape fraction that we derive from our Monte-Carlo code agrees very well with this analytic result for zero and non-zero dust albedos \citep[as was also found by other authors, e.g.][]{Laursen07,Forero11,Yajima11}. 

\section{Derivation of Analytic Expressions for the Surface Brightness \& Polarization Profiles}
\label{app:an}
%

\begin{figure}
\vbox{\centerline{\epsfig{file=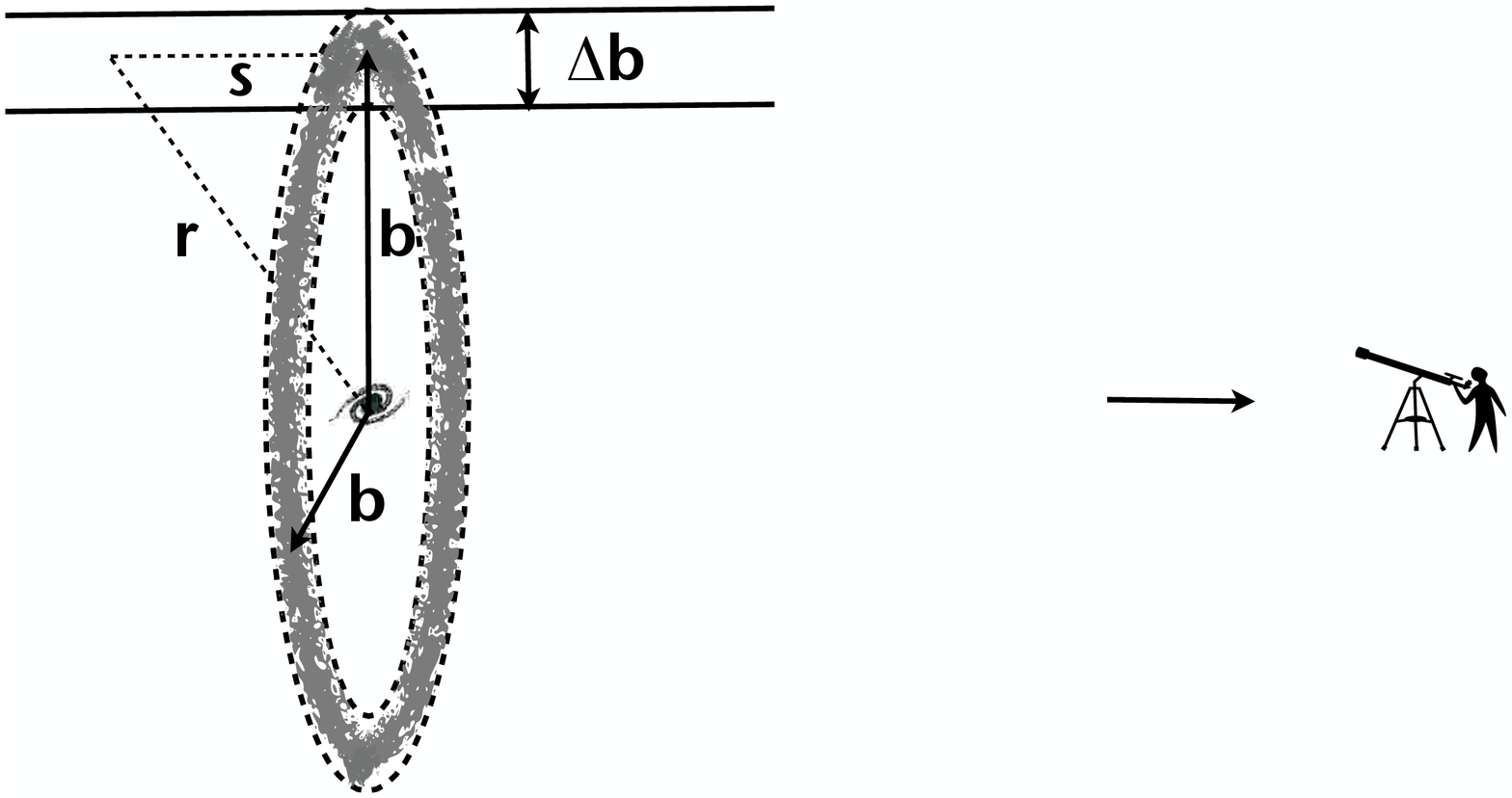,angle=90,width=8.5truecm}}}
\vspace{0mm}
\caption[]{This figure depicts our adopted geometry and coordinates
  for the analytic calculation of the surface brightness profile of
  the scattered radiation.}  
\label{fig:b}
\end{figure}  

The surface brightness (per unit area) of scattered radiation at
impact parameter $b \pm \Delta b/2$ is given by
\begin{eqnarray}
S(b)=\frac{1}{2 \pi b \Delta b}\int_{-\infty}^{\infty}ds \hs 2 \pi b
\int_{-\infty}^{\infty}dx  \times \nonumber \\ \times f_{\rm
  esc}(x'',b,s) s_{\rm in}(s,b,x)p_{\rm scat}(s,b,x) \Delta b.
\label{eq:b}
\end{eqnarray}
The factor $2 \pi b \hs ds \hs s_{\rm in}(s,b,x) $ in the first line
denotes the total flux incident on a ring of radius $b$ at frequency
$x$ at line of sight coordinate $s$. The probability $p_{\rm
  scat}(s,b,x)\Delta b$ denotes the fraction of this flux that is
scattered towards the observer. The probability $p_{\rm
  scat}(b,s,x)\Delta b=f_{\rm c}(r) \Delta b\frac{r}{b} \times
(1-\exp[-\tau_{\rm clump}(x',s,b)])$, where the first term denotes the
probability that the photon hits a clump over the path of length
$\Delta r = \Delta b \frac{r}{b}$, and the second term denotes the
probability that the photon gets scattered by this particular
clump. The incoming flux at location $(b,s)$ and frequency $x$, as
well as the expression for  $\exp[-\tau_{\rm clump}(x',s,b)]$, is
given in the main paper. The term  $f_{\rm esc}(x'',b,s)$ denotes the
fraction of the photons scattered at location $(b,s)$ and frequency
$x$, that are observed. Finally, in the main paper we expressed impact
parameter in proper kpc, and we added a factor (kpc/asec)$^{2}$ to
convert the surface brightness profile into the proper units of erg
s$^{-1}$ cm$^{-2}$ arcsec$^{-2}$. 

Equation~\ref{eq:b} is relevant for the total observed flux. We can
derive the expressions for {\it polarized} fluxes $S_l(b)$ and
$S_r(b)$ from the scattering matrix that describes scattering in the
wing of the line, which is given by \citep[][and references
  therein]{DL08}
\begin{equation}
R= \frac{3}{2}\left( \begin{array}{cc} \mu^2 & 0 \\ 0 & 1 \end{array}
\right)
\label{eq:matrix}
\end{equation} where the scattering matrix is defined as

\begin{equation}
\left( \begin{array}{c} I_l' \\ I_r' \end{array}
\right)=R\left( \begin{array}{c} I_l \\ I_r \end{array} \right).
\end{equation} Here, $I_l'$ and $I_r'$ denote the components of the scattered intensity parallel and perpendicular to the plane of scattering, respectively. The incoming flux is directly coming directly from the source and is therefore likely unpolarized, in which case we have $I_l=I_r=\frac{1}{2}I$. We can therefore write
\begin{eqnarray}
\left.\begin{array}{ll} S_l(b)\\ S_r(b)
     \end{array}
\right\} =\frac{3}{4}\times \frac{1}{2 \pi b \Delta
  b}\int_{-\infty}^{\infty}ds \hs 2 \pi b \int_{-\infty}^{\infty}dx
\times \nonumber \\ \times f_{\rm esc}(x'',b,s) s_{\rm
  in}(s,b,x)p_{\rm scat}(s,b,x) \Delta b\times
\left\{\begin{array}{ll} \mu^2\\ 1
     \end{array}.
\right.
\label{eq:b}
\end{eqnarray}

\section{Directional-Dependent Redistribution}
 \label{app:redist}
 
The frequency redistribution function, often denoted with $R(x_{\rm
  out}|x_{\rm in})$, denotes the probability density function for the
photon frequency $x_{\rm out}$ after scattering, given it had a
frequency $x_{\rm in}$ prior to scattering. This function is an
important quantity in the Ly$\alpha$ radiative transfer process, and
analytic expressions have been known for decades \citep[see e.g.][and
  references therein]{Lee74}. The frequency redistribution function
averages over all possible scattering angels, $\mu$. However, the
redistribution function varies strongly with $\mu$: this is most
evident when considering the case $\mu=1$. Here, energy conservation
implies that the photon frequency before and after scattering must be
identical, and hence $R(x_{\rm out}|x_{\rm in},\mu=1) \neq R(x_{\rm
  out}|x_{\rm in})$. In this section we present a complete derivation
of $R(x_{\rm out}|x_{\rm in},\mu)$.
 
The photon frequencies before and after scattering are
related\footnote{This assumes coherence in the frame of the atom,
  which is relevant at the densities and temperatures of interest (see
  e.g. Dijkstra et al. 2006 for a more detailed discussion).}  through
the angle at which the photon is scattered, and the total 3-D velocity
of the hydrogen atom that scatters the photon as \citep[e.g.][]{D06}

\begin{equation}
x_{\rm out}=x_{\rm in}-\frac{{\bf v}\cdot {\bf k}_{\rm in}}{v_{\rm
    th}}+\frac{{\bf v}\cdot {\bf k}_{\rm out}}{v_{\rm
    th}}+g(\mu-1)+\mathcal{O}(v^2_{\rm th}/c^2),
\label{eq:xoxi1}
\end{equation} where $g=h\Delta\nu_{\alpha}/(2k_BT_{\rm c})=2.6\times 10^{-4}(T_{\rm c}/10^4\hs{\rm K})^{-1/2}$ is the fractional amount of energy that is transferred per scattering event \citep{Field59}. Throughout this calculation we safely ignore recoil \citep{Adams71}. 

For simplicity, but without loss of generality, we can define a
coordinate system such that ${\bf k}_{\rm in}=(1,0,0)$, and ${\bf
  k}_{\rm out}=(\mu,\sqrt{1-\mu^2},0)$, i.e. the photon wavevectors
lie entirely in the x-y plane. Following \citet{Ahn00} we decompose
the atom's velocity into components parallel  ($v_{||}$) and
orthogonal ($v_y$ and $v_z$) to ${\bf k}_{\rm in}$, and we have ${\bf
  v}=(v_{||},v_y,v_z)$.  Eq~\ref{eq:xoxi1} can then be recast as

\begin{eqnarray}
\label{eq:xoxi2}
x_{\rm out}=x_{\rm in}-\frac{v_{||}}{v_{\rm
    th}}+\frac{v_{||}\mu}{v_{\rm th}}+\frac{v_y\sqrt{1-\mu^2}}{v_{\rm
    th}}\equiv \\ \nonumber x_{\rm in}-u+u\mu+w\sqrt{1-\mu^2},
\end{eqnarray} where we have introduced the dimensionless velocity parameters $u=v_{||}/v_{\rm th}$ and $w=v_{y}/v_{\rm th}$. Note that the value of $v_z$ is irrelevant in this equation.\\

The velocities $u$ and $w$ are unrelated, and the most general way of
writing the directional redistribution function is
\begin{eqnarray}
R(x_{\rm out}|\mu,x_{\rm in})=\mathcal{N}\int_{-\infty}^{\infty} du
\int_{-\infty}^{\infty} dw \\ \nonumber R(x_{\rm out}|\mu,x_{\rm
  in},u,w)P(u|\mu,x_{\rm in})P(w|\mu, x_{\rm in}),
\end{eqnarray} where $\mathcal{N}$ denotes the normalization constant. The integral over $w$ can be eliminated by utilising Eq~\ref{eq:xoxi2}. That is, we replace $R(x_{\rm out}|\mu,x_{\rm in},u,w)=\delta_D(f[w_{\rm u}])=\delta_D(w-w_{\rm u})/\frac{df}{dw}$, in which $f[w_{\rm u}]=x_{\rm out}-x_{\rm in}+u-u\mu-w_{\rm u}\sqrt{1-\mu^2}$. We find

\begin{equation}
R(x_{\rm out}|\mu,x_{\rm in})=\mathcal{N}\int_{-\infty}^{\infty}  du
P(u|\mu,x_{\rm in})P(w_u|\mu, x_{\rm in}),
\label{eq:R2b}
\end{equation} where the factor $1/\frac{df}{dw}=1/\sqrt{1-\mu^2}$ is absorbed by the normalization constant $\mathcal{N}$.\\

The conditional absorption probabilities for both $w$ and $u$ cannot
depend on the subsequent emission direction, and therefore
$P(u|\mu,x_{\rm in})=P(u|x_{\rm in})$ and $P(w_u|\mu, x_{\rm
  in})=P(w_u|x_{\rm in})$. Furthermore, the conditional PDF for $w$
does not depend on $x_{\rm in}$ either. This is because $w$ denotes
the normalized velocity of the scattering atom in a direction
perpendicular to ${\bf k}_{\rm in}$, and the frequency that the atoms
`sees' does not depend on $w$. The absorption probability can
therefore not depend on $w$, and $P(w_u|x_{\rm
  in})=P(w_u)=\sqrt{\frac{m_p}{2\pi k_BT}}\exp(-w_u^2)$, where we
assumed a Maxwell-Boltzmann distribution for the atoms' velocities.

The expression for $P(u|x_{\rm in})$ can be obtained from Bayes
theorem \citep[see e.g.][]{Lee74}, which states that $P(u|x_{\rm
  in})=P(u,x_{\rm in})/P(x_{\rm in})=P(x_{\rm in}|u)P(u)/P(x_{\rm
  in})$, in which $P(x_{\rm in}|u)$ denotes the absorption probability
for a single atom that has a speed $u$, and $P(x_{\rm in}) \propto
\sigma(x_{\rm in})$. Therefore, $P(u|x_{\rm in})\propto P(x_{\rm
  in}|u) P(u)/\sigma(x_{\rm in})$, where $P(x_{\rm
  in}|u)=\frac{3\lambda_{\alpha}^2}{8\pi}\frac{A_{\alpha}^2}{[\omega_{\alpha}(x_{\rm
      in}-u)v_{\rm th}/c]^2+A_{\alpha}^2/4}$ \citep[e.g.][]{RL79}. If
we substitute this into Eq~\ref{eq:R2b} and absorb all factors that
can be pulled out of the integral into the normalization constant
$\mathcal{N}$, then we get

\begin{eqnarray}
R(x_{\rm out}|\mu,x_{\rm in})=\mathcal{N}\int_{-\infty}^{\infty} du
\frac{ \exp(-u^2)}{(x_{\rm in}-u)^2+a^2_v}  \times \\ \nonumber \times
     {\rm exp}\Big{[}-\Big{(}\frac{\Delta x+u(\mu-1)}{\sqrt{1-\mu^2}}
       \Big{)}^2\Big{]}
\end{eqnarray}, where we introduced $\Delta x\equiv x_{\rm in}-x_{\rm out}$. The normalization constant can be computed analytically, and we have

\begin{eqnarray}
R(x_{\rm out}|\mu,x_{\rm in})=\frac{a_v}{\pi^{\frac{3}{2}}\phi(x_{\rm
    in)}\sqrt{1-\mu^2}} \int_{-\infty}^{\infty} du
\frac{\exp(-u^2)}{(x_{\rm in}-u)^2+a_v^2} \\ \nonumber \times {\rm
  exp}\Big{[}-\Big{(}\frac{\Delta x+u(\mu-1)}{\sqrt{1-\mu^2}}
  \Big{)}^2\Big{]}\hspace{5mm}{\rm for}\hs|\mu| <1.
\label{eq:rxmu}
\end{eqnarray} For $\mu=1$ we have $x_{\rm out}=x_{\rm in}$ (Eq~\ref{eq:xoxi2}). For $\mu =-1$ we have $x_{\rm out}=x_{\rm in}-2u$, and  we have
$R(x_{\rm out}|\mu=-1,x_{\rm in})=\frac{1}{2}P(u_{\rm c}|x_{\rm in})$,
in which $u_{\rm c}=(x_{\rm in}-x_{\rm out})/2$. We have verified that
these analytic expressions are in excellent agreement with results
obtained from Monte-carlo calculations.

\end{document}